\documentclass{iopjournal_justified_3}

\usepackage{bm}
\usepackage{graphicx}
\usepackage{color}
\usepackage{hyperref}
\usepackage{latexsym}
\usepackage{amsthm}
\usepackage{amssymb}
\usepackage{amsmath}
\usepackage{cite}

\usepackage{bm}
\usepackage{xcolor}

\def\bea{\begin{eqnarray}}
\def\eea{\end{eqnarray}}

\usepackage{microtype}
\begin{document}
\articletype{Paper} 

\title{Topology and higher-order global  synchronization \\on directed and hollow simplicial and cell complexes}

\author{Runyue Wang$^{1,*}$\orcid{0009-0005-9386-1655}, Timoteo Carletti$^{2,*}$\orcid{0000-0003-2596-4503} and Ginestra Bianconi$^{1,*}$\orcid{0000-0002-3380-887X}\hfill}

\affil{$^1$ Centre for Complex Systems, School of Mathematical Sciences, Queen Mary University of London, Mile End Road, London, E1 4NS, UK \hfill}

\affil{$^2$ Department of Mathematics \& naXys, Namur Institute for Complex Systems, Université de Namur, Rue Grafé 2, Namur, 5000, Belgium \hfill}

\affil{$^*$Authors to whom any correspondence should be addressed.\hfill}

\email{runyue.wang@qmul.ac.uk,timoteo.carletti@unamur.be,ginestra.bianconi@gmail.com \hfill}

\keywords{Higher-order networks, Topological higher-order dynamics, Global Topological Synchronization\hfill}

\begin{abstract}
Higher-order networks encode  the many-body interactions of complex systems ranging from the brain to biological transportation networks. Simplicial and cell complexes are ideal higher-order network representations for investigating  higher-order topological dynamics where dynamical variables are not only associated with nodes, but also with edges, triangles, and higher-order simplices and cells. Global Topological Synchronization (GTS) refers to the dynamical state in which identical oscillators associated with higher-dimensional simplices and cells oscillate in unison. On standard unweighted and undirected complexes this dynamical state can be achieved only under strict topological and combinatorial conditions on the underlying discrete support. In this work we consider  generalized higher-order network representations  including directed  and hollow complexes.  Based on an in depth  investigation of  their topology defined by  their associated algebraic topology operators and Betti numbers, we determine under which conditions  GTS can be observed. We show that directed  complexes always admit a global topological synchronization state independently of their topology and structure. However, we demonstrate that for directed  complexes  this dynamical state cannot be asymptotically stable. While hollow  complexes require more stringent topological conditions to sustain global topological synchronization,  these topologies can favor both the existence and the stability of global topological synchronization with respect to undirected and unweighted complexes.
\end{abstract}

\section{Introduction}
Higher-order networks are increasingly recognized as fundamental representations of complex systems as they offer new theoretical paradigms to encode the many-body interactions between two or more nodes~\cite{bick2023higher,bianconi2021higher,battiston2020networks,salnikov2018simplicial,boccaletti2023structure}. 
These generalized network structures have been shown to affect dynamical processes \cite{millan2025topology,majhi2022dynamics} unfolding on them, including synchronization~\cite{millan2020explosive,carletti2023global,gallo2022synchronization,gambuzza2021stability,skardal2019abrupt,skardal2020higher,carletti2020dynamical,majhi2025patterns,pal2024global,anwar2024collective}, diffusion~\cite{torres2020simplicial,krishnagopal2023topology,taylor2015topological}, random walks~\cite{carletti2020random,schaub2020random}, and percolation \cite{sun2021higher,sun2023dynamic,bianconi2023theory}.
Among the different representations of higher-order networks~\cite{torres2021and}, simplicial and cell complexes  are the most suitable to investigate the interplay between topology and dynamics \cite{millan2025topology} which has profound effects in the collective behavior of complex systems, in their design, processing and control with dynamical systems and AI algorithms. Examples of fundamental dynamical systems in which topology shapes dynamics include the topological Dirac equation \cite{bianconi2021topological}, topological diffusion \cite{schaub2020random,torres2020simplicial}, topological  pattern formation \cite{giambagli2022diffusion,muolo2024three,muolo2024turing} as well as Topological Synchronization~\cite{millan2020explosive,ghorbanchian2021higher,nurisso2023unified,zaid2025designing,arnaudon2022connecting,bavcic2025phase,ziegler2026hodge} and Global Topological  Synchronization \cite{carletti2023global,carletti2025global,wang2024global}. A common feature of these proposals is that they go beyond the node-centered approach in which dynamical variables are exclusively assigned to the nodes of the higher-order structure. Specifically, topological, higher-order dynamics embraces a topological approach and describes the  dynamical state of a higher-order network by associating dynamical variables not only with its nodes, but also with its edges, triangles, and higher-order simplices and cells. This dynamical description of higher-order networks is attracting significant attention in particular in neuroscience applications where topological investigation of structure and dynamics \cite{petri2014homological,reimann2017cliques,santoro2023higher,faskowitz2022edges,santos2023emergence} has been shown to reveal features that are not observable with more traditional statistical approaches.

However, simplicial and cell complexes have the limitation that the simplices and cells from which they are built describe undirected interactions with a trivial topology. Since in neuroscience, biological transportation networks and other applications of topological higher-order dynamics, directionality is of paramount importance, it is also fundamental to go beyond the undirected simplicial complex framework. Moreover, the deviation of the building blocks of higher-order networks from structures with a trivial topology can also provide further flexibility that can be used in the study of dynamical processes and AI algorithms to process higher-order topological dynamics.

Important progress in understanding synchronization on directed hypergraphs has already been made in the context of node-based dynamics \cite{gallo2022synchronization}. However, investigating how directionality affects higher-order topological dynamics on simplicial and cell complexes is still in its infancy~\cite{schaub2020random,arnaudon2022connecting,Dorchain2024,guo2025topological}. It is well known that formulating a well-defined topological framework that encodes directionality is particularly challenging. Various approaches have been proposed so far \cite{gong2024higher,schaub2020random,grigor2020path,suwayyid2024persistent,mulas2020coupled,jost2019hypergraph}.
Among these approaches, we mention in particular directed simplicial complexes in which the simplices are associated with a single orientation. These directed simplicial and cell complexes provide a very well-defined way to generalize homology in the directed setting and have been used to study higher-order topological dynamics (random walks and topological, higher-order Kuramoto-like models) in Refs.\cite{schaub2020random,arnaudon2022connecting}. Another relevant generalization of more standard simplicial and cell complex representations are hollow complexes proposed in the context of image processing in Ref. \cite{sardellitti2024topological} which are not compact and include a hollow in their center.

In this work, we investigate Global Topological Synchronization (GTS) \cite{carletti2023global} on directed and hollow simplicial and cell complexes shedding light on the interplay between their generalized topology and dynamical processes.
On standard simplicial and cell complexes, GTS is a dynamical state that occurs only under strict conditions on their topology. Specifically, a global topological synchronized state involving the $n$-cochain, i.e., the dynamical signal associated to $n$-dimensional simplices or cells, occurs only if the Hodge Laplacian admits in its kernel a vector having elements of constant absolute value.
While some cell complexes, like the square lattice tessellation of a torus, admit a global synchronized state for simplices of every dimension, on simplicial complexes it can be shown that this condition implies that odd dimensional signals can never globally synchronize.
In a previous work~\cite{wang2024global} we have shown that by considering weighted homology~\cite{baccini2022weighted} allows edge signals to globally synchronize on weighted simplicial complexes, provided that the weights are chosen appropriately and the vector having elements of constant absolute value is divergence-free.

In this work we discuss  how directed and hollow simplicial complexes affect global topological synchronization. The topology encoded  by the generalized algebraic topology operators (boundary operators and Hodge Laplacian) of such structures is here derived, allowing us to address naturally the dynamical properties of GTS. In particular, we will focus not only on the existence of such a state but also on its stability, which is essential to observe GTS in real systems. For simplicity, throughout this work, we will consider unweighted complexes which will allow us to single out the effect of directionality and adopting hollow cells with non-trivial topology instead of standard cells and simplices.

Specifically, in this work, we will show that directed (single orientation) simplicial and cell complexes always allow a global topological synchronized state for any topological signal, regardless of its dimension. This scenario is akin to node-based synchronization in simple networks. However, we observe a significant difference with node-based dynamics, as in directed simplicial complexes GTS cannot be asymptotically stable, reflecting a topological aspect of GTS which goes beyond the standard Master-Stability-Function (MSF) approach originally defined  for node-based dynamics \cite{fujisaka1983stability,pecora1998master,boccaletti2006complex}.
Additionally, we reveal how hollow simplicial complexes, favour the existence of a global topological synchronized state with respect to standard simplicial complexes.  We reveal that on hollow simplicial complexes there are more restrictions for the existence of GTS with respect to directed single orientation simplicial complexes, however, some topologies can sustain asymptotically stable GTS. To illustrate the scenario in a simple setting, edge signals can globally synchronize only if there exists a vector having elements of constant absolute value on each edge of the simplicial complex which is divergence-free. This occurs, for instance, on a $2D$ torus tessellated by hollow triangles. 
To complete our investigation of generalized complexes we compare the hollow simplicial complexes with cell complexes that tessellate the same structure. Interestingly, we demonstrate that global topological synchronization cannot be observed in the latter, even if it can be observed in the hollow simplicial complexes, thus revealing the unique dynamical implications of choosing the hollow simplicial complex representation. Finally, we briefly discuss how these results generalize to directed and hollow cell complexes.

\section{Global Topological Synchronization on standard simplicial and cell complexes}
\subsection{Higher-order topological dynamics on simplicial and cell complexes}
Simplicial complexes \cite{bianconi2021higher} are set of simplices closed under the inclusion of their faces. Therefore simplicial complexes include nodes ($0$-simplices), edges ($1$-simplices), triangles ($2$-simplices), tetrahedra ($3$-simplices) and so on. 
A $n$-simplex $\alpha$ is a set of $n+1$ nodes, i.e.
\bea
\alpha=[i_0,i_1,\ldots,i_n].
\eea
Typically simplices are associated to a double orientation, so that 
\bea
[i_0,i_1,\ldots,i_n]=(-1)^{\tau(\pi)}[i_{\pi(0)},i_{\pi(1)},\ldots,i_{\pi(n)}]
\eea
where $\pi$ is an arbitrary permutation of the indices and $\tau(\pi)$ is its associated parity. In particular, here and in the following we will assume, unless stated otherwise, that the $n$-simplices with $n>0$ have a positive orientation when the labels of the nodes are listed in increasing order or on any even permutation of this order.
The closure requirement implies that if a  $n$-dimensional simplex belongs to a simplicial complex $\mathcal{K}$ then also all the $m$-dimensional simplices with $m<n$ constructed starting from proper subset of its nodes belong to $\mathcal{K}$.
Cell complexes are natural generalization of simplicial complexes where the building blocks are regular polytopes, including all $2$-dimensional polygons, the $3$-dimensional Platonic solids, and in higher dimensions simplices, orthoplexes and hypercubes.
We indicate with $D$ the dimension of the simplicial and cell complexes under consideration, which is given by the maximum dimension of any of their cells.
Investigating how the properties of AI algorithms and higher-order dynamical processes change when the original dynamics defined on a simplicial complex is investigated on a more general  cell complexes is a hot topic in the interdisciplinar field focusing on topological higher-order dynamics and  topological AI \cite{carletti2023global,carletti2025global,sardellitti2024topological,bavcic2025phase}.
In particular, early on in this young research field, it was discussed (see Ref. \cite{carletti2023global}) that topological global synchronization can be favoured by the adoption of cell complexes as we briefly summarize in the following paragraph.

\subsection{Higher-order topological dynamics}
Global topological synchronization is a higher-order dynamical process for topological signals, i.e., dynamical variables associated with higher-order dimensional simplices and cells.
Let us consider a general simplicial (or cell) complex having $N_{[n]}$ simplices (or cells) of dimension $n$.
A $n$-dimensional topological signal  ${\phi}$ is treated as a $n$-cochain ${\phi} \in {C}^n$ and can be represented by a vector $\bm\phi\in \mathbb{R}^{N_{[n]}\times d}$  in which every element $\phi_{\alpha}\in \mathbb{R}^d$ represents the dynamical variable associated with the corresponding simplex (cell) $\alpha$.
The $n$-cochain can describe, for instance, currents associated with each edge of the network. Thus, if the simplices have double orientation, the sign that a cochain acquires on a given simplex depends on the choice adopted for the positive orientation of the simplex.
Among the other things, this implies that if an edge is positively oriented from node $i$ to node $j$, a positive current goes from $i$ to $j$ while if the orientation is reverted, the same current will acquire a negative sign.

Topological signals can be treated with topological operators, by including boundary and coboundary operators, Hodge Laplacians \cite{grady2010discrete,horak2013spectra,lim2020hodge} and the Dirac operator \cite{bianconi2021topological}. Of special interest in this work are the $n$-order Hodge Laplacians ${\bf L}_{[n]}$ that  determine the higher-order topological diffusion among $n$-simplices (or cells) passing through $(n+1)$-dimensional and $(n-1)$-dimensional simplices (or cells).
The Hodge Laplacians ${\bf L}_{[n]}$ can thus be decomposed into the sum of the up- and down- Hodge Laplacians ${\bf L}_{[n]}^{up}$, ${\bf L}_{[n]}^{down}$, i.e.
\bea
{\bf L}_{[n]}={\bf L}_{[n]}^{up}+{\bf L}_{[n]}^{down}.
\eea
For simplices and cells having double orientation, these latter operators can be defined by means of the boundary matrices ${\bf B}_{[n]}$ which are $N_{[n-1]}\times N_{[n]}$ matrices whose elements are given by
\begin{equation}
[{\bf B}_{[n]}]_{\alpha\alpha^{\prime}}=\left\{\begin{array}{cc}-1 &\mbox{if} \ \alpha\not\sim \alpha^{\prime}, \\1 &\mbox{if}\  \alpha\sim \alpha^{\prime},\\
0 & \mbox{otherwise},\end{array}\right.\label{boundary}
\end{equation}
where $\sim$, $\not\sim$ indicate whether the $(n-1)$ simplex $\alpha$ and the $n$ simplex $\alpha^{\prime}$ have coherent or incoherent orientation respectively.
Specifically, we have 
\bea
{\bf L}_{[n]}^{up}={\bf B}_{[n+1]}[{\bf B}_{[n+1]}]^{\top}\quad \text{and}\quad {\bf L}_{[n]}^{down}=[{\bf B}_{[n]}]^{\top}{\bf B}_{[n]}.
\eea
where since the {\it boundary of the boundary is null}, the boundary operators satisfy the fundamental topological relation 
\bea
{\bf B}_{[n]}{\bf B}_{[n+1]}={\bf 0}.
\eea
The Hodge Laplacians play a fundamental role in algebraic topology \cite{grady2010discrete}. Here we briefly mention few fundamental results  relevant for the following discussion.
First of all, the dimension of the kernel of the Hodge Laplacian is a topological invariant. Specifically, it is equal to the $n$-Betti number, $\beta_n$, counting the $n$-dimensional holes of the complex, i.e.,
\bea
\mbox{dim}(\mbox{ker}({\bf L}_{[n]}))=\beta_n.
\eea
Secondly, any $n$ cochain can be decomposed uniquely thanks to Hodge decomposition 
\bea
C^n=\mbox{im}([{\bf B}_{[n]}]^{\top})\oplus\mbox{ker}({\bf L}_{[n]})\oplus\mbox{im}({\bf B}_{[n+1]}).
\eea
This implies, for instance, that an edge signal can be decomposed uniquely into a gradient flow, an harmonic signal and a curl flow.
Finally, one of the consequences of Hodge decomposition is that the up- and down- Hodge Laplacians ${\bf L}_{[n]}^{up}$ and ${\bf L}_{[n]}^{down}$ not only commute, but obey also 
\bea
{\bf L}_{[n]}^{up}{\bf L}_{[n]}^{down}={\bf 0},\quad {\bf L}_{[n]}^{down}{\bf L}_{[n]}^{up}={\bf 0}.
\eea
This implies that  
\bea
\mbox{ker}({\bf L}_{[n]}^{up})\supseteq\mbox{im}({\bf L}_{[n]}^{down}),\quad \mbox{ker}({\bf L}_{[n]}^{down})\supseteq\mbox{im}({\bf L}_{[n]}^{up}).
\eea
\subsection{Global Topological Synchronization (GTS)}
Here we treat GTS describing a dynamical state in which a $n$-cochain acquires the same dynamics modulus an overall possible change of sign on each $n$-simplex (or $n$-cell) of the complex.
We assume that each $n$-simplex ($n$-cell) supports an identical oscillator, and that the neighbor $n$-simplices ($n$-cells) are interacting through a diffusive coupling enforced by the $n$-dimensional Hodge Laplacian ${\bf L}_{[n]}$.
Under these assumptions, the higher-order topological dynamics of each $n$-simplex ($n$-cell) obeys:
\bea
\frac{d\bm{\phi}_\alpha}{dt} = F(\bm{\phi}_\alpha) - \sigma \sum_{\alpha' \in \mathcal{Q}_n}[\bm{L}_{[n]}]_{\alpha,\alpha'} h(\bm{\phi}_{\alpha'}), \label{GTS_dyn}
\eea 
where $\sigma$ is the coupling constant, and $\mathcal{Q}_n$ represents the set of all $n$-dimensional simplices (or  cells) in the complex.
Moreover, as discussed in Ref.~\cite{carletti2020dynamical}, where this dynamics is proposed for simplicial and cell complexes with double orientation, in order to guarantee that $\bm\phi$ is a proper cochain, whose elements change sign under change of orientation of each simplex, we must impose that $F,h: \mathbb{C}^d \to \mathbb{C}^d$ are odd functions.  
 
GTS is achieved when all oscillators  follow  the same dynamics after a transient evolution, modulo a  possible difference in the  sign  i.e., 
\bea
\bm{\phi}(t) = {\bf w}(t)\otimes {\bf u},
\eea  with 
\bea
\frac{d{{\bf w}(t)}}{dt} = F({\bf w}(t)),\eea
and ${\bf u}\in \mathbb{R}^{N_{[n]}}$ being a vector of elements with unitary absolute value, i.e. \bea |u_{\alpha}|=1,\; \forall \alpha.\eea 
By requiring that the dynamical system of equations $(\ref{GTS_dyn})$ admits this GTS state, we obtain the spectral properties that the complex must have to sustain this dynamics.
This implies that the diffusive coupling must leave the dynamics unaffected, and this occurs if and only if ${\bf u}$ is in the kernel of the Hodge Laplacian, i.e.
\bea
\bm{L}_{[n]} {\bf u} = \bm{0}.\label{GTS_Laplacian}
\eea
In Ref.\cite{carletti2020dynamical} the topological conditions for GTS to occur on standard simplicial and cell complexes  where simplices and cells have double orientations have been identified and discussed.
Specifically, thanks to Hodge decomposition, the condition $(\ref{GTS_Laplacian})$ implies that  the GTS state can be achieved only if ${\bf u}$ is both in the kernel of ${\bf L}_{[n]}^{up}$ and ${\bf L}_{[n]}^{down}$ this implies that the signal ${\bf u}$ that is constant in modulus in each $n$-simplex (or cell) should obey at the same time 
\bea
{\bf B}_{[n+1]}^{\top}{\bf u}={\bf 0},\quad {\bf B}_{[n]}{\bf u}={\bf 0}.\label{GTS_Boundary}
\eea
Thus, this implies that if ${\bf u}$ is an edge signal, it should be at the same time divergence-free and curl-free.
If the GTS state exists, an important problem is to establish its stability. According to the MSF approach, we can establish the stability of the GTS by investigating the dynamical properties of the linearized dynamics. Indicating with $\delta \bm\phi$ the small perturbations with respect to the GTS state, we obtain that these small perturbations obey
 \bea
 \hspace{15mm}\frac{d\delta{\bm \phi}}{dt}&=&\left[ {{\bf I}_{N_n}\otimes {\bf {J}}}_{F}-\sigma {{\bf L}}_{[n]}\otimes {\bf {J}}_{h}\right]\delta{\bm\phi}\, ,\label{stability}
    \eea
where we have indicated with  ${{\bf {J}}}_{F}$ and ${{\bf {J}}}_{h}$ the Jacobians of $F$ and $h$ evaluated on the GTS state.
The GTS  will be stable  if and only if the maximum Lyapunov exponent of the system in Eqs.(\ref{stability}) is negative.

These results indicate that GTS is significantly different from topological synchronization of node signals when complexes are formed by simplices and cell with double orientation.
In particular, while for node signals, global synchronization is always a possible dynamical state, and can be observed on any network provided that it is also stable, GTS can be observed only on topologies that satisfy Eq.(\ref{GTS_Laplacian}) when the simplices and cells have double orientation.
Interestingly, this implies that cell complexes can favour the existence of a GTS state. In fact, on (unweighted) simplicial complexes, odd dimensional signals can never satisfy the first of equations (\ref{GTS_Boundary}) and thus can never globally synchronize. Instead, the combinatorics of cell complexes is such that both conditions expressed by Eqs. (\ref{GTS_Boundary}) can be realized in specific cell complexes.
In particular $D$ dimensional tori tessellated by hypercube admit GTS for topological signals of any dimension.
Note, however, that, as we will discuss later in more detail, when  the Hodge Laplacian ${\bf L}_{[n]}$ admits more than one eigenvector of type ${\bf u}$ in its kernel (scenario realized, for instance, for the $D$-dimensional torus)  GTS aligned along one of these eigenvectors is only marginally stable with respect to perturbations aligned with respect to the other ones, regardless of the choices of the dynamical parameters.

These results open new perspectives in GTS and naturally lead to the following research question: what are the conditions under which a stable GTS state exists on generalized simplicial and cell complexes?
While in Ref.~\cite{wang2024global} we have investigated this question for weighted simplicial complexes, here we focus on other generalized complexes, including directed simplicial complexes where simplices have a single orientation, hollow simplicial complexes and together with directed and hollow cell complexes.

\section{Generalized complexes and their topology}
\begin{figure}
    \centering
    \includegraphics[width=0.8\linewidth]{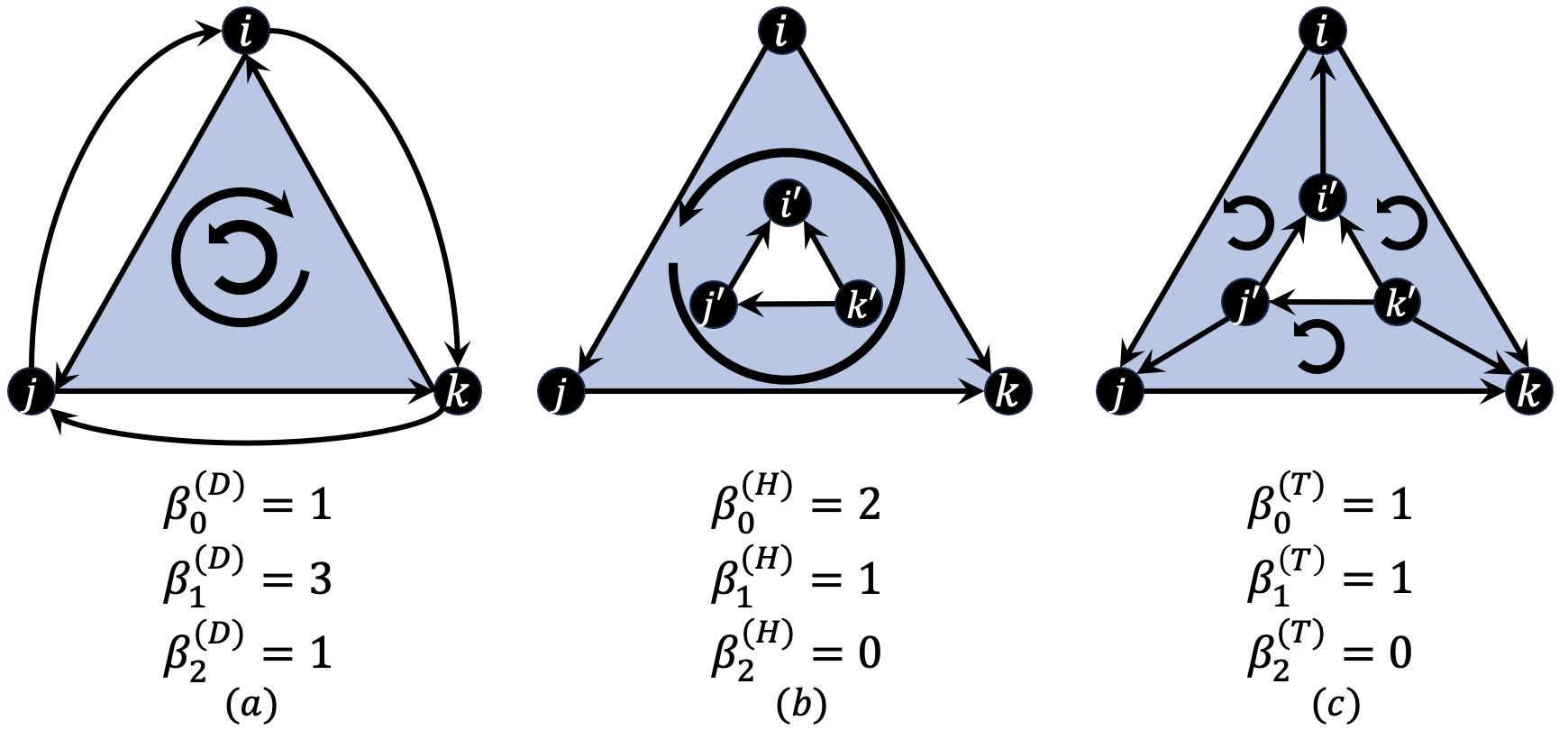}
    \caption{Schematic illustration of the directed simplicial complex (DSC), hollow simplicial complex (HSC) and tessellated hollow simplicial complex (TSCC). Panel (a) shows a DSC composed of simplices (nodes, edges, and triangles) having single orientation. Note that for simplicity we illustrate only a  triangle with single orientation, but the DSC includes two single oriented triangles with opposite orientation. Panel (b) is a HSC, and panel (c) is a THSC induced from the HSC shown in panel (b). Below each of these generalized complexes we indicate their corresponding Betti numbers following Eqs.(\ref{betti_D}), (\ref{betti_H}),(\ref{betti_T}) respectively.}
    \label{fig1}
\end{figure}

In this work, we consider generalized complexes by including the directed simplicial complex (DSC) where simplices have single orientations, hollow simplicial complexes (HSC) where the simplices do not have a trivial topology and include an hollow in their center, and the tessellated hollow simplicial  complexes (THSC) that tessellate the HSC. These generalized complexes are illustrated in Figure \ref{fig1}, where we focus on $2$-dimensional complexes formed by nodes, edges, and faces. 
As we will see in the following, the properties of identical oscillators coupled by diffusive dynamics change significantly among these generalized complexes. Therefore our results will highlight the important role that adopting these generalized topological structures have on the conclusion drawn in Ref.\cite{carletti2023global} summarized in the previous section.

\subsection{Directed Simplicial Complexes (DSC)}

 The directed simplicial complex (DSC) \cite{schaub2020random,arnaudon2022connecting} are formed by a set of simplices closed under the inclusion of their faces. The difference from standard simplicial complexes is that in the DSC each simplex $\alpha$ of dimension $n>0$ is associated to two  single orientation simplices while the nodes have only positive orientation as in the standard case.  If for simplicity we consider the cases of $1$-simplices,  for each edge $(i,j)$ in DSC we consider two linearly independent $1$-simplices: $[i,j]$ and $[j,i]$. 
This implies that the number of directed $n$-dimensional simplices $N_{[n]}^{(D)}$ is given by 
\bea
N_{[0]}^{(D)}&=&N_{[0]},\nonumber \\
N_{[n]}^{(D)}&=&2N_{[n]}\quad \mbox{for}\ 0<n\leq D.
\eea
It follows that  the directed Euler characteristic $\chi^{(D)}$ is related to the Euler characteristic of the undirected network $\chi$  by 
\bea
\chi^{(D)}\equiv\sum_{n=0}^D (-1)^n N_{[n]}^{(D)}=2\chi-N_{[0]}.
\eea
Their associated boundary matrices ${\bf B}^{(D)}_{[n]}$ are $N_{[n-1]}^{(D)}\times N_{[n]}^{(D)}$ matrices for DSC and can be constructed from the corresponding boundary operators ${\bf B}_{[n]}$ for standard simplicial complexes as
\bea
{\bf B}^{(D)}_{[1]} &=& {\bf B}_{[1]} \mathcal{I}_{[1]}, \nonumber \\
{\bf B}^{(D)}_{[n]} &=& \mathcal{I}^\top_{[n-1]} {\bf B}_{[n]} \mathcal{I}_{[n]},
\eea
where $\mathcal{I}_{[n]}$ are the lifting matrices
\bea
\mathcal{I}_{[n]} = \left( \bm{I}_{N_{[n]}},\; -\bm{I}_{N_{[n]}} \right),
\eea
where here and in the following $\bm{I}_{N_{[n]}}$ indicates the $N_{[n]}\times N_{[n]}$ identify matrix.
For instance, a full directed triangle, as shown in Fig $\ref{fig1}$ panel (a), has directed boundary matrices ${\bf B}_{[1]}^{(D)}$ and ${\bf B}_{[2]}^{(D)}$ given by 
\bea
{\bf B}_{[1]}^{(D)}=\left[\begin{array}{c|cccccc}
&{[i,j]}&{[j,k]}&{[i,k]}&{[j,i]}&{[k,j]}&{[k,i]}\\
\hline
{[i]}&-1&0&-1&1&0&1\\
{[j]}&1&-1&0&-1&1&0\\
{[k]}&0&1&1&0&-1&-1
\end{array}\right],\quad  {\bf B}_{[2]}^{(D)}=\left[\begin{array}{c|cc}
&{[i,j,k]}&{[k,j,i]}\\
\hline
{[i,j]}&1 &-1\\
{[j,k]}&1&-1\\
{[i,k]}&-1&1\\
{[j,i]}&-1&1\\
{[k,j]}&-1&1\\
{[k,i]}&1&-1\\
\end{array}\right].
\eea
The directed Hodge  Laplacians ${\bf L}_{[n]}^{(D)}$ are constructed from the directed boundary operators. Specifically ${\bf L}_{[n]}^{(D)} $ are $N_{[n]}^{(D)}\times N_{[n]}^{(D)}$ matrices defined as 
\bea
{\bf L}^{(D)}_{[0]} &=& {\bf L}^{(D),up}_{[0]}, \nonumber \\
{\bf L}^{(D)}_{[n]} &=& {\bf L}^{(D),up}_{[n]}+{\bf L}^{(D),down}_{[n]},
\eea
with  the up and down directed Hodge  Laplacians ${\bf L}^{(D),up}_{[n]}$, and ${\bf L}^{(D),down}_{[n]}$ given by 
\bea
{\bf L}^{(D),up}_{[n]}&=&{\bf B}^{(D)}_{[n+1]} {\bf B}^{(D),\top}_{[n+1]}\nonumber \\
{\bf L}^{(D),down}_{[n]}&=&{\bf B}^{(D),\top}_{[n]} {\bf B}^{(D)}_{[n]}.
\eea
A direct inspection reveals that the up and down directed Hodge  Laplacians obey Hodge decomposition as 
\bea
{\bf L}^{(D),up}_{[n]}{\bf L}^{(D),down}_{[n]}={\bf 0},\quad \mbox{and}\quad {\bf L}^{(D),down}_{[n]}{\bf L}^{(D),up}_{[n]}={\bf 0}.
\eea
We observe that by considering DSC, where simplices have single orientation, has a drastic effect on the relation between the spectral properties of the Hodge Laplacian and the underlying topology of the complexes.
In particular, as shown in Appendix \ref{ApA} we notice that the dimension of the kernel of the directed Hodge Laplacians ${\bf L}_{[n]}^{(D)}$ for the DSC, defining the directed Betti numbers $\beta_n^{(D)}$ is much bigger than the corresponding Betti number $\beta_n$ of the standard simplicial complex for $n>0$. For $n=0$ we have the classic result
\bea
\mbox{dim}(\mbox{ker}({\bf L}_{[0]}^{(D)}))&\equiv&\beta_0^{(D)}=\beta_{0},\nonumber \\
\mbox{dim}(\mbox{ker}({\bf L}_{[n]}^{(D)}))&\equiv&\beta_n^{(D)}=N_{[n]}+\beta_{n}, \quad \mbox{for}\ 0<n\leq D.
\label{betti_D}
\eea

Taking as an example the directed $1$-Hodge Laplacian for the DSC shown in Figure $\ref{fig1}(a)$, although the original simplicial complex (the triangle) has Betti number $\beta_1=0$, the Betti number $\beta_1^{(D)}$ of the DSC is given by  $\beta_1^{(D)}=3$. Intuitively, the Betti number $\beta_1^{(D)}$ counts as true the ``holes"  formed by traversing each edge back and forth along the two corresponding $1$-simplices with single orientation. 

Interestingly, we found the consistent topological relation between the alternating sum of the directed Betti numbers and the directed Euler characteristic, i.e.
\bea
\chi^{(D)}=\sum_{n=0}^D(-1)^n\beta_n^{(D)}.
\eea

\subsection{Hollow Simplicial Complexes (HSC)}
The hollow simplicial complexes (HSC) proposed in Ref. \cite{sardellitti2024topological} have  a more restrictive structure and are formed by simplices having double orientation as in the case of standard simplicial complexes. The hollow simplicial complexes of dimension $D\geq 2$ are formed by hollow simplices of dimension $D$ also called hollow facets. For instance if we consider a $2D$ HSC the triangles of the complex will be hollow, while if we consider $3D$ HSC the tetrahedra of the complex will be hollow etc.
By considering the illustrative example of a simple hollow triangle shown in Figure $\ref{fig1}(b)$ we can intuitively describe its construction starting from the directed triangle shown in Figure $\ref{fig1}(a)$ where we double the nodes $i,j,k$ with $i<j<k$ by generating the ``internal replica nodes" $i^{\prime},j^{\prime}, k^{\prime}$ labelled according to the numerical order $i^{\prime}>j^{\prime}>k^{\prime}$ forming an internal hollow triangle. Finally, for  the hollow triangle  we choose as positive orientation the one defined by $[i,j,k;i^{\prime},j^{\prime},k^{\prime}]$. Therefore  instead of considering, accordingly to the DSC construction, for each edge $(i,j)$ two single orientation simplices $[i,j]$ and $[j,i]$, we consider the two simplices $[i,j]$ and $[j^{\prime},i^{\prime}]$ having double orientation.
In this way, it is easy to show that the hollow boundary matrices ${\bf B}^{(H)}_{[1]},{\bf B}^{(H)}_{[2]}$ for the hollow triangle become 
\bea
{\bf B}_{[1]}^{(H)}=\left[\begin{array}{c|cccccc}
&{[i,j]} &{[j,k]} &{[i,k]}& {[j^{\prime}, i^{\prime}]} & {[k^{\prime}, j^{\prime}]} & {[k^{\prime}, i^{\prime}]}\\
\hline
{[i]}&-1&0&-1&0&0&0\\
{[j]}&1&-1&0&0&0&0\\
{[k]}&0&1&1&0&0&0\\
{[i^{\prime}]} &0&0&0&1&0&1\\
{[j^{\prime}]} &0&0&0&-1&1&0\\
{[k^{\prime}]} &0&0&0&0&-1&-1
\end{array}\right],\quad  {\bf B}_{[2]}^{(H)}=\left[\begin{array}{c|cc}
&[i,j,k;i^{\prime},j^{\prime},k^{\prime}] \\
\hline
{[i,j]}&1 \\
{[j,k]}&1\\
{[i,k]}&-1\\
{[j^{\prime},i^{\prime}]} &-1\\
{[k^{\prime},j^{\prime}]} &-1\\
{[k^{\prime},i^{\prime}]} &1\\
\end{array}\right].\label{Htriangle}
\eea
This construction can be easily generalized to define the  $D$-dimensional HSC. Starting from a $D$-dimensional pure undirected simplicial complex (i.e., formed by $D$-dimensional undirected simplices and their faces) for any $D$-dimensional simplex $\alpha=[i_0,i_1,\ldots, i_D]$ with $i_0<i_1<\ldots < i_D$ we generate a replica of their nodes $i_0^{(\alpha)},i_1^{(\alpha)}\ldots, i_D^{(\alpha)}$ with a label chosen such that $i_0^{(\alpha)}>i_1^{(\alpha)}>\ldots> i_D^{(\alpha)}$ which we place in the interior of the simplex and will represent the nodes defining its internal hole. The HSC will include all the nodes of the original undirected simplicial complex and all the  replica nodes corresponding to each $D$ simplices. The $D$-dimensional simplices of the original undirected simplicial complex become the double oriented hollow facets of the HSC. The HSC will also include all the original $0<n<D$ simplices of the undirected simplicial complex plus all the set of possible simplices of dimension $0<n<D$ which can be defined among each set of replica nodes, each taken with double orientation. 
Finally, for  the generic hollow $D$ simplex $\alpha$  we choose as positive orientation the one defined by $[i_0,i_1,\ldots,i_D;i_0^{(\alpha)},i_1^{(\alpha)},\ldots,i_D^{(\alpha)}]$.
It therefore follows that the number of $n$-dimensional simplices $N_{[n]}^{(H)}$ of the HSC is given by 
\bea
N_{[n]}^{(H)}&=&N_{[n]}+\binom{D+1}{n+1}N_{[D]},\quad \mbox{for}\  0\leq n\leq D-1,\nonumber \\
N_{[D]}^{(H)}&=&N_{[D]}.
\eea
Thus the hollow Euler characteristic $\chi^{(H)}$ is related to the Euler characteristic of the undirected network $\chi$  by 
\bea
\chi^{(H)}\equiv\sum_{n=0}^D (-1)^n N_{[n]}^{(H)}=\chi+\Big(1+(-1)^{D-1}\Big)N_{[D]}.
\eea
Let us adopt  as a the basis for the $0\leq n\leq D$  simplices, the list of $n$-simplices in which we order first the  (positively oriented) simplices present also in the original simplicial complex, and then the (positively oriented) internal replica simplices corresponding to each original simple $\alpha$, following the same order as for the corresponding external simplices, as illustrated in the specific case of a single triangle in Eq.(\ref{Htriangle}). 
In this case, the boundary matrices ${\bf B}_{[n]}^{(H)}$ of hollow simplicial complexes have dimensions $N_{[n-1]}^{(H)}\times N_{[n]}^{(H)}$. For $n< D$ these hollow boundary matrices have the $(1+N_{[D]})\times (1+N_{[D]})$ block structure
\bea
{\bf B}_{[n]}^{(H)}=\begin{pmatrix}{\bf B}_{[n]}&0 &0&\ldots &0\\
0&\hat{\bf B}_{[n]}&0&\ldots& 0\\
0&0&\hat{\bf B}_{[n]}&\ldots& 0\\
\vdots&\vdots&\vdots &\vdots&\vdots\\
0&0&0&\ldots&\hat{\bf B}_{[n]}\end{pmatrix},
\eea
where $\hat{\bf B}_{[n]}$ is the $n$-th boundary matrix (with $0<n<D$) associated with a single generic undirected $D$ internal simplex, written in the basis of positively oriented internal simplices,  and having elements given by Eq.(\ref{boundary}).
Instead, the boundary matrix ${\bf B}_{[D]}^{(H)}$ has a different block structure, as the hollow $D$-dimensional simplices connect the  replicated $D-1$ faces with the $D$ dimensional faces.
In particular,  the boundary matrix ${\bf B}_{[D]}^{(H)}$ has the block structure of a $(1+N_{[D-1]})$-dimensional column vector given by
\bea
{\bf B}_{[D]}^{(H)}=\begin{pmatrix}{\bf B}_{[D]}\\
\hat{\bf B}_{[D]}^{(1)}\\
\hat{\bf B}_{[D]}^{(2)}\\
\vdots\\
\hat{\bf B}_{[D]}^{(r)}\\
\vdots\\
\hat{\bf B}_{[D]}^{({N_{[D]}})}
\end{pmatrix}.
\eea
With our choice of the basis for the $(D-1)$-dimensional simplices, $\hat{\bf B}_{[D]}^{(r)}$ indicates the $D\times N_{{D}}$ boundary matrix obtained from ${\bf B}_{[D]}$ by retaining only the $r$-th column relative to the $D$ simplex $\alpha_r$ (of basis column vector ${\bf e}_{\alpha_r}$) multiplied by minus one, i.e.
\bea
\hat{\bf B}_{[D]}^{(r)}=-{\bf B}_{[D]}{\bf e}_{\alpha_r}{\bf e}_{\alpha_r}^{\top}.\label{boundary_r}\eea 
and then removing all the null rows of this matrix that do not correspond to the internal replica nodes of simplex $\alpha_r$.
The hollow Hodge Laplacian ${\bf L}_{[n]}^{(H)}$ are built from the hollow boundary operator as usual and thus they are defined as 
\bea
{\bf L}^{(H)}_{[0]} &=& {\bf L}^{(H),up}_{[0]}, \nonumber \\
{\bf L}^{(H)}_{[n]} &=& {\bf L}^{(H),up}_{[n]}+{\bf L}^{(H),down}_{[n]},
\label{LH}
\eea
with  the up and down hollow Hodge  Laplacians ${\bf L}^{(H),up}_{[n]}$, and ${\bf L}^{(H),down}_{[n]}$ given by 
\bea
{\bf L}^{(H),up}_{[n]}&=&{\bf B}^{(H)}_{[n+1]} {\bf B}^{(H),\top}_{[n+1]}\nonumber \\
{\bf L}^{(H),down}_{[n]}&=&{\bf B}^{(H),\top}_{[n]} {\bf B}^{(H)}_{[n]}.
\eea
Also in this case, a direct inspection reveals that the up and down hollow Hodge  Laplacians obey Hodge decomposition as 
\bea
{\bf L}^{(H),up}_{[n]}{\bf L}^{(H),down}_{[n]}={\bf 0},\quad \mbox{and}\quad {\bf L}^{(H),down}_{[n]}{\bf L}^{(H),up}_{[n]}={\bf 0}.
\eea
Since the simplices belonging to each internal holes define empty $D$-dimensional simplices and are connected to the rest of the HSC only by the $D$-dimensional hollow facet, it is possible (see Appendix \ref{ApA2} for details) to compute  the dimension of the kernel of the hollow Hodge Laplacian ${\bf L}_{[n]}^{(H)}$ defining the hollow Betti numbers $\beta_n^{(H)}$. 
Specifically, the hollow Betti numbers $\beta_n^{(H)}$ and the original Betti numbers $\beta_n$ of the corresponding standard simplicial complex are related by the following relations: 
\bea
\mbox{dim}(\mbox{ker}({\bf L}_{[0]}^{(H)}))&\equiv&\beta_0^{(H)}=\beta_{0}+N_{[D]},\nonumber \\
\mbox{dim}(\mbox{ker}({\bf L}_{[n]}^{(H)}))&\equiv&\beta_n^{(H)}=\beta_{n},\; \; \mbox{for}\; 0<n<D-1,\nonumber \\
\mbox{dim}(\mbox{ker}({\bf L}_{[D-1]}^{(H)}))&\equiv&\beta_{D-1}^{(H)}=\beta_{D-1}-\beta_{D}+N_{[D]}, \nonumber \\
\mbox{dim}(\mbox{ker}({\bf L}_{[D]}^{(H)}))&\equiv&\beta_D^{(H)}=0.
\label{betti_H}
\eea
\begin{figure}
    \centering
    \includegraphics[width=0.8\linewidth]{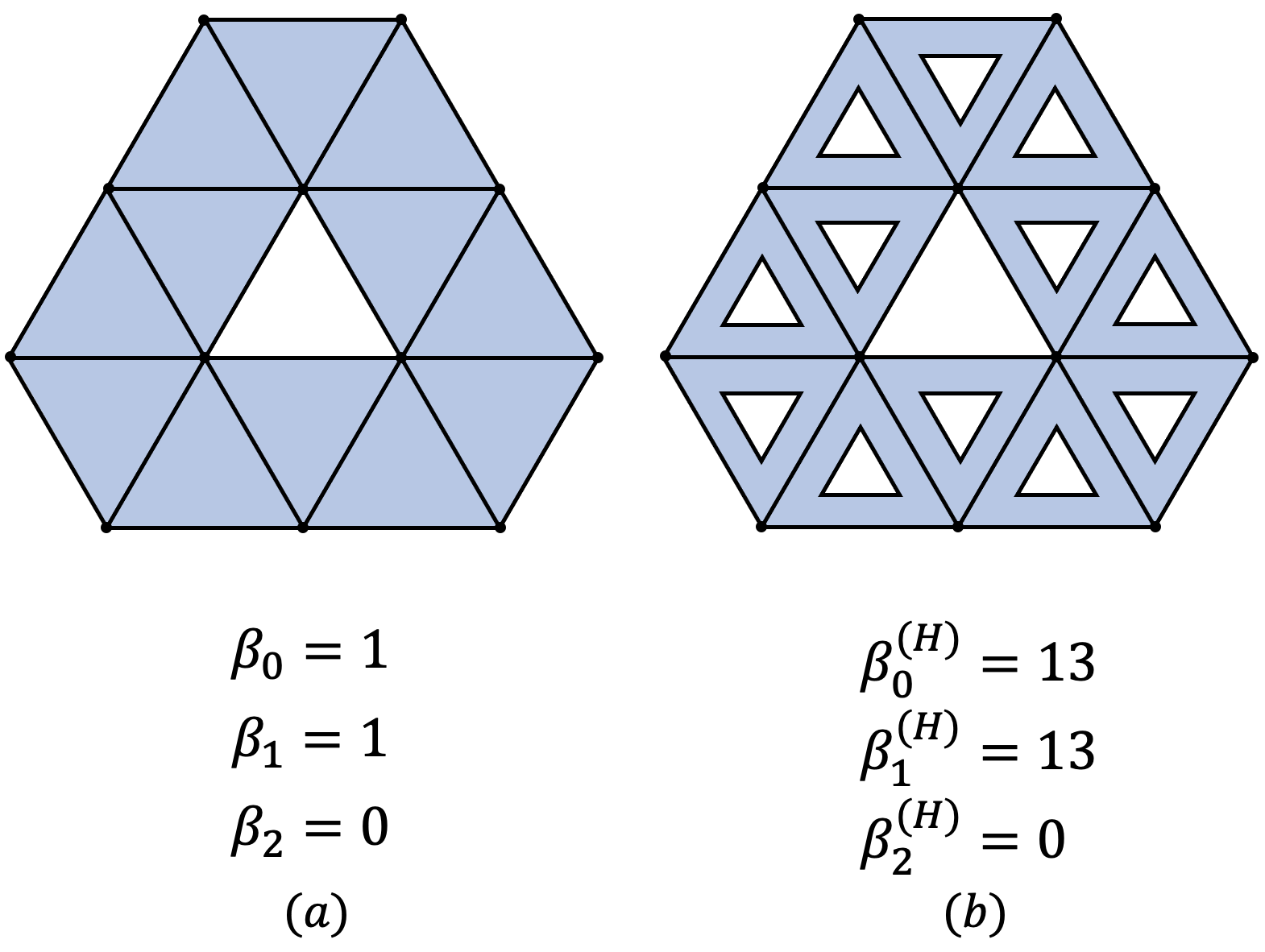}
    \caption{Illustration of the topology of the HSC as captured by their associated Betti number $\beta^{(H)}$ given by Eqs.(\ref{betti_H}) whose derivation is given in \ref{ApA2}. Panel (a) shows an original pure $D=2$ dimensional simplicial complex, panel (b) shows associated HSC. The  Betti numbers $\beta_n$ is the original network (panel (a)) and the Betti numbers $\beta_n^{(H)}$ of the HSC (panel (b)) are indicated below the corresponding simplicial complex visualizations.}
    \label{fig:ApB}
\end{figure}
In other words, if $\beta_D=0$ the dimension  of the kernel of the $D-1$-dimensional hollow Hodge Laplacian will count in addition to the original $(D-1)$-dimensional holes of the standard simplicial complex, also the $(D-1)$-dimensional holes induced by the hollow in the middle of each $D$-dimensional hollow simplex. When $\beta_D\neq0$ the hollow Betti number ${\beta}^{(H)}_{D-1}$ will be reduced  exactly by $\beta_D$. Moreover, since also the skeleton of the HSC includes an additional disconnected component for each hollow simplex, the dimension of the kernel of ${\bf L}_{[0]}^{(H)}$ will increase by one unit for each $D$-dimensional hollow simplex. The dimension of the kernel of the hollow Hodge Laplacians ${\bf L}_{[n]}^{(H)}$ with $0<n<D$ will remain unchanged. While the dimension of kernel of the $D$ Hodge Laplacian will be zero, independently on the Betti number $\beta_{D}$ of the original undirected simplicial complex.
Also in this case,  we found the consistent topological relation between the alternating sum of the hollow Betti numbers and the hollow Euler characteristic, i.e.
\bea
\chi^{(H)}=\sum_{n=0}^D(-1)^n\beta_n^{(H)}.
\eea
As an illustration of these results, in Figure $\ref{fig:ApB}$ we show how the Betti number of an original simplicial complex change when we generate a HSC out of it.

\begin{figure}
    \centering
    \includegraphics[width=0.8\linewidth]{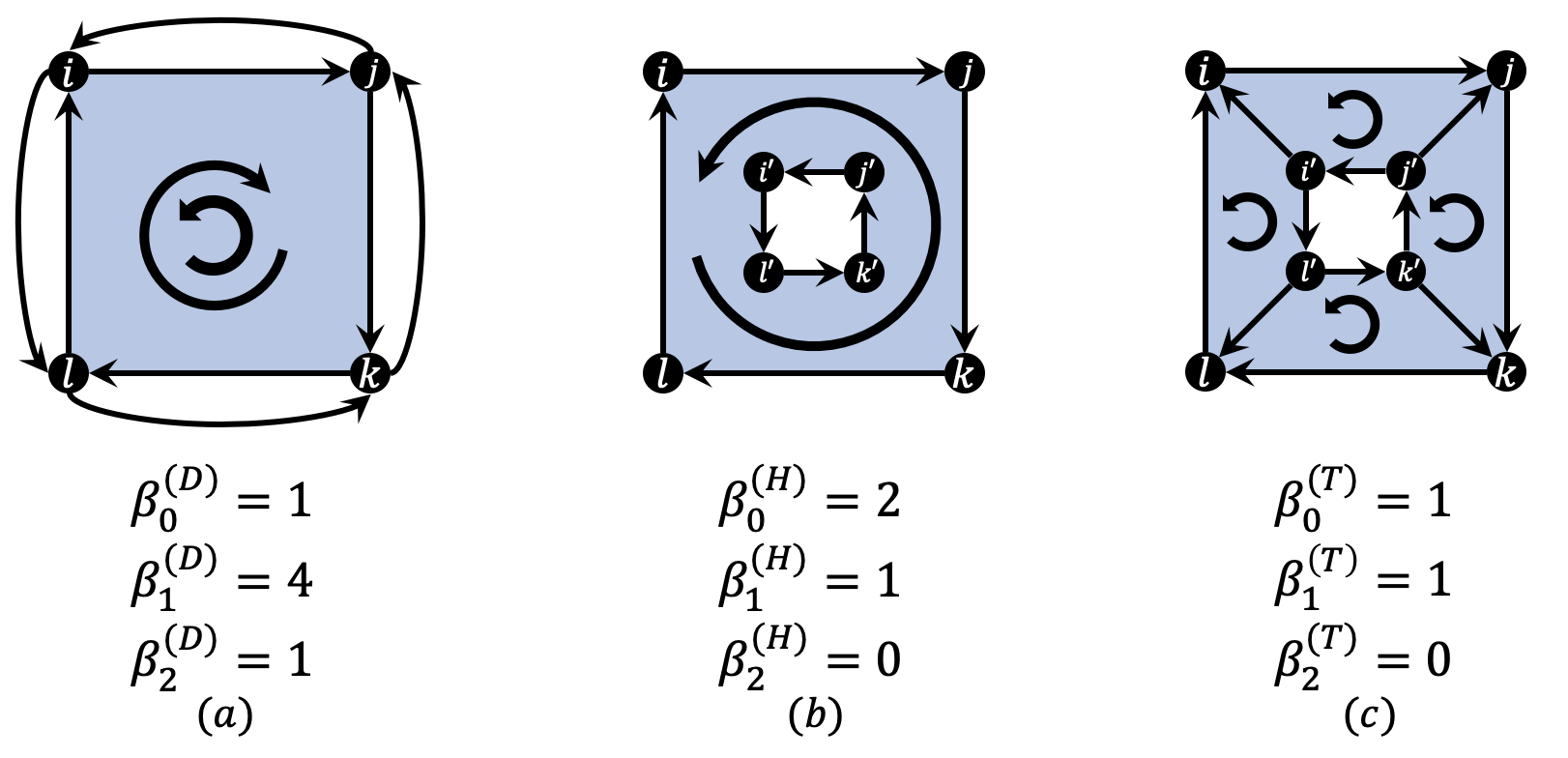}
    \caption{Schematic illustration of three types cell complexes which generalize the DSC, HSC and THSC shown in Figure $\ref{fig1}$. Panel (a) shows a directed cell complex (DCC) composed of faces, (nodes, edges and squares) having single orientation. Note that for simplicity we illustrate only a  square with single orientation, but the DSC includes two single oriented squares with opposite orientation. Panel (b) is a hollow cell complex (HCC), and panel (c) is a tesselated hollow cell complex (THCC) induced from the HCC shown in panel (b). Below each of these generalized complexes we indicate their corresponding Betti numbers following Eqs.(\ref{betti_D}), (\ref{betti_H}),(\ref{betti_T}) respectively.}
    \label{fig1b}
\end{figure}
\subsection{Tessellated Hollow Simplicial Complexes (THSC)}
In our investigation of generalized complexes we finally consider a standard cell complex which tessellates the HSC described above. We call such cell complexes   {\it tessellated hollow simplicial  complexes} (THSC). The THSCs have the same number of nodes of the corresponding HSCs but include more cells of dimensions $0<n\leq D$. This is illustrated for the case of a THSCs constructed from a hollow triangle in Figure $\ref{fig1}(c)$ which include more $2$-dimensional and $1$-dimensional cells of the HSC. 
Since THSC are standard cell complexes, they are associated with the standard notion of Hodge Laplacian. However, as the THSCs have $N_{[D]}$ additional holes with respect to the original simplicial complex,  their associated Betti numbers $\beta_n^{(T)}$ are related to the Betti number $\beta_n$ of the original undirected simplicial complex as
\bea
\mbox{dim}(\mbox{ker}({\bf L}_{[n]}))&\equiv&\beta_n^{(T)}=\beta_{n},\; \; \mbox{for}\; 0\leq n<D-1, \nonumber \\
\mbox{dim}(\mbox{ker}({\bf L}_{[D-1]}))&\equiv&\beta_{D-1}^{(T)}=\beta_{D-1}+N_{[D]},\; \; \nonumber \\ 
\mbox{dim}(\mbox{ker}({\bf L}_{[D]}))&\equiv&\beta_D^{(T)}=0.\; \;  
\label{betti_T}
\eea
Note that with respect to the HSC the dimension of the kernel of the graph Laplacian ${\bf L}_{[0]}$ is simply given by the Betti number $\beta_0$ of the original simplicial complex. This is due to the fact that replica nodes in the THSC are connected to the original nodes of each simplex by construction.
Since the THSC is a particular example of a cell complex, the usual properties of homology and cohomology apply, thus the Laplacians obey Hodge decomposition and the alternating sum of the Betti numbers $\beta_n^{(T)}$ is equal to the Euler characteristic $\chi^{(T)}$ of the THCC which is related to the original undirected simplicial complex by 
\bea
\chi^{(T)}=\chi+(-1)^{D-1}(N_{[D]}+\beta_{D}). 
\eea
\subsection{Directed, Hollow and Tessellated Cell Complexes}
The DSC, HSC, and THSC can be directly extended to cell complexes leading to Directed Cell Complexes (DCC), Hollow Cell Complexes (HCC), and Tessellated Hollow Cell Complexes  (THCC) that are schematically represented in Figure $\ref{fig1b}$.
The algebraic topology is directly derived from the one of the previously discussed simplicial complexes, where the algebraic topology operators, instead of being formulated in terms of the boundary operators of the original undirected simplicial complex, are expressed in terms of the oundary operators of the original cell complex. Therefore, the Betti numbers of DCC, HCC and THCC naturally obey Eqs.(\ref{betti_D}) (\ref{betti_H}) and (\ref{betti_T}) respectively.
\section{Global topological synchronization on generalized complexes}
\subsection{Existence and stability of the GTS state}
We model the topological dynamics of higher-order identical oscillators described in Eq.(\ref{GTS_dyn}) by adopting specifically  the Stuart-Landau (SL) model with Hodge-Laplacian couplings. The SL model is the normal form for a supercritical Hopf bifurcation that captures the dynamics of a stable nonlinear oscillation. The $n$-dimensional topological  signal $\bm{\phi}$ on each face $\alpha$ represents an identical  oscillator $\phi_{\alpha} \in \mathbb{C}$. The uncoupled time evolution is given by \bea
\frac{d{\phi}_\alpha}{dt}=F(\phi_{\alpha}) = \delta \phi_{\alpha} - \mu|\phi_{\alpha}|^2\phi_{\alpha}\eea with $\delta,\mu \in \mathbb{C}$ given by $\delta=\delta_\Re+\textrm{i}\delta_\Im,\mu=\mu_\Re+\textrm{i}\mu_\Im$. The coupling between oscillators is given by the nonlinear function $h(\phi_{\alpha}) = \phi_{\alpha} |\phi_{\alpha}|^{m-1}$ with $m \in \mathbb{N}$. To ensure that the  uncoupled dynamics   admits a stable limit cycle of amplitude $\rho_0$ and frequency $\omega$, given  
\bea
\phi_{\alpha}=\hat{z}(t)=\rho_0e^{\textrm{i}\omega t}  \ \mbox{with}\ 
\rho_0 = \sqrt{\frac{\delta_\Re}{\mu_\Re}},\quad
\omega=\delta_\Im-\delta_\Re\frac{\mu_\Im}{\mu_\Re}
\label{z}
\eea we assume 
\bea
\delta_\Re > 0, \quad \mu_\Re > 0.
\label{SLstab}
\eea
By including a diffusive coupling the resulting dynamics on generalized complexes reads
\bea
\frac{d{\phi}_\alpha}{dt} = \delta {\phi}_\alpha - \mu|{\phi}_\alpha|^2{\phi}_\alpha - \sigma \sum_{\alpha' \in \mathcal{Q}_n}[{\bf L}_{[n]}^{(G)}]_{\alpha,\alpha'} {\phi}_{\alpha'} |{\phi}_{\alpha'}|^{m-1},
\label{SL}
\eea
where $\sigma=\sigma_\Re+\textrm{i}\sigma_\Im\in \mathbb{C}$ and ${\bf L}_{[n]}^{(G)}$ indicates the directed Hodge Laplacian ${\bf L}_{[n]}^{(D)}$ on DSC, the hollow Hodge Laplacian ${\bf L}_{[n]}^{(H)}$ on HSC and the standard Hodge Laplacian ${\bf L}_{[n]}$ on the THCC.
The GTS state occurs when all the oscillators undergo the same dynamics, i.e
\bea
\phi_{\alpha}=\rho_{\alpha}e^{\textrm{i}\theta_{\alpha}}=w(t)\; \; \ \forall \alpha\in Q_n,
\eea
where $\rho_{\alpha},\theta_{\alpha}$ indicate respectively  the polar and the angular coordinates of the complex valued edge signal ${\phi_\alpha}$.
The  generalized Kuramoto order parameter $R_2$  that is not sensible to phase differences of $\pi$ among the oscillators and is defined as
\bea
R_2=\frac{1}{N_{[n]}}\left|\sum_{\alpha\in \mathcal{Q}_{n}}\frac{\phi_{\alpha}^2}{\rho_0^2}\right|=\frac{1}{N_{[n]}}\left|\sum_{\alpha\in \mathcal{Q}_{n}}\frac{\rho_{\alpha}^2}{\rho_0^2}e^{\textrm{i}2\theta_{\alpha}}\right|.
\eea
Thus the onset of the GTS state is indicated by the generalized Kuramoto order parameter asymptotically acquiring a stationary value $R_2=1$.

A straightforward generalization of Eq.(\ref{GTS_Laplacian}) leads to the necessary and sufficient condition for the existence of the GTS state on generalized complexes, which reads
\bea
\bm{L}_{[n]}^{(G)} {\bf u} = \bm{0},\label{gen_GTS_Laplacian}
\eea
where ${\bf u}\in \mathbb{R}^{N_{[n]}}$ has elements $u_{\alpha}$ of unitary absolute value, i.e. $|u_{\alpha}|=1$. 
Condition which implies 
\bea
{\bf B}_{[n+1]}^{(G),\top}{\bf u}={\bf 0},\quad {\bf B}_{[n]}^{(G)}{\bf u}={\bf 0}.\label{gen_GTS_Boundary}
\eea
where ${\bf B}^{(G)}_{[n]}$ indicates the generalized boundary matrix, given by the directed (${\bf B}^{(D)}_{[n]}$), hollow (${\bf B}^{(H)}_{[n]}$) or standard boundary matrix (${\bf B}_{[n]}$) for DSC, HSC, and THSC respectively.
While these are the essential conditions on the complex topology to ensure the existence of GTS, this state can only be observed if the dynamical system in Eq.(\ref{SL}) is also dynamically stable. The MSF approach determines whether this dynamical state is stable with respect to perturbations perpendicular to the kernel of ${\bf L}_{[n]}$. By indicating with $\Lambda_{[n]}^{(k)}$ the $k$-th eigenvalue of the generalized Laplacian matrix, the MSF requires that the  largest real part of the Floquet eigenvalues $\lambda_k$ determining the dynamics associated with the $k$-th eigenmode of the deviations away from the synchronized solution is negative (see Appendix \ref{ApB} for details).
However, for establishing the stability of the GTS state we should also take care of the perturbations aligned to different modes lying on the kernel of the Hodge Laplacians ${\bf L}_{[n]}$. To this end we distinguish two scenarios: in the first scenario there is only a single eigenvector of type ${\bf u}$ in the kernel of ${\bf L}_{[n]}$, in the second scenario there is more than one eigenvector of type ${\bf u}$ in the kernel of ${\bf L}_{[n]}$. In the first case the stability of  the GTS is ensured by Eqs.(\ref{SLstab}), together with the usual MSF characterization, while in the second case the different eigenvectors of the Hodge Laplacian kernel which are of  type ${\bf u}$ will all define neutral stability eigenmodes independently on the result of the MSF analysis. 

\begin{figure}
    \centering
    \includegraphics[width=0.8\linewidth]{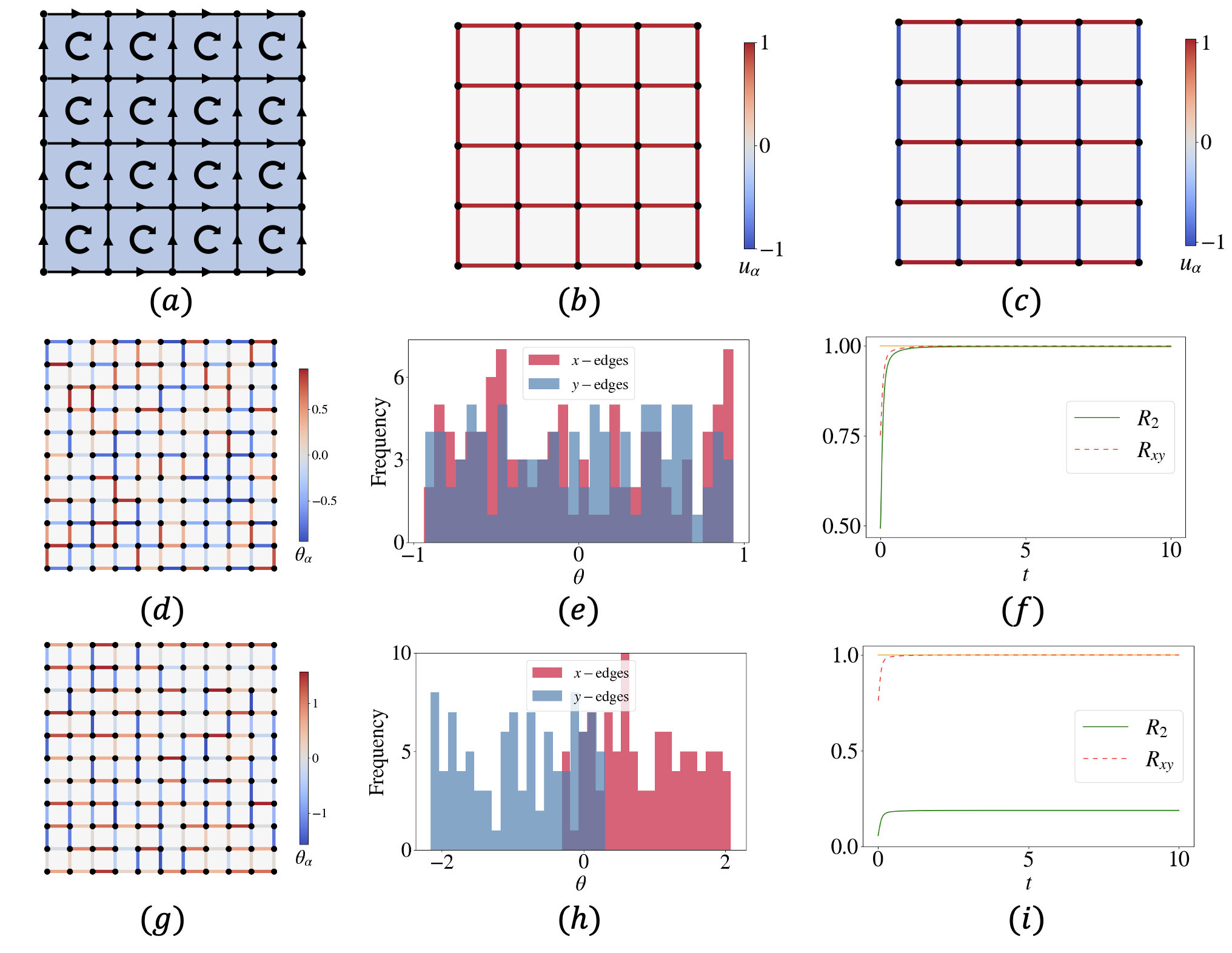}
    \caption{Harmonic eigenstates of the square lattice tessellation of a $2D$ torus and stability of the GTS defined on it. 
    Panels (a)-(c) represent respectively the chosen orientation of the faces of the $2D$ torus (panel (a)) and the two orthogonal eigenvectors of type ${\bf u}$ in the kernel of the Hodge Laplacian ${\bf L}_{[1]}$. Panel (b) represents the eigenvector with elements defined on  both $x$-edges and $y$-edges  equal to one. Panel (c) represents the eigenvector in which elements defined on the $x$-edges are equal to one, while elements define on $y$-edges are equal to minus one.
    Panels (d)-(f) demonstrate that  GTS can be achieved ($R_2$ converges to $1$)  when the random initial condition  is unbiased on both $x$ and $y$ links. This is due to the fact that in this case  the dynamics, to a very good approximation, includes only perturbations orthogonal to the kernel of ${\bf L}_{[1]}$.  Panels (g)-(i) demonstrate that instead when the initial condition on  the edge signal is biased ($R_2$ converges to $1$) the neutral stability of GTS becomes apparent and GTS cannot be achieved. However our results reveal that in both cases the topological oscillators associated to $x$-edges and the one associated to $y$-edges do synchronize among themselves as indicated by $R_{xy}$ converging to $1$ in both cases. The phase $\theta_{\alpha}$ associated with the complex valued topological signal $\phi_{\alpha}$ defined on each edge $\alpha$ is plotted on each edge of the torus for the unbiased (panel (d)) and the biased  (panel (g)) initial condition. The corresponding distributions of the phases $\theta_{\alpha}$ associated to $x$-type edges and $y$-type edges are shown in panels (e) and (h). The resulting dynamics of the generalized order parameters {$R_2$  and $R_{xy}$} are shown as a function of time $t$ in panels (f) and (i).The considered dynamics is given by the SL model with parameters $\delta=40+4.3\textrm{i},\mu=40+40.1\textrm{i},\sigma=1+1.1\textrm{i},m=3$.}
    \label{fig_cell0}
\end{figure}

As an illustrative case, let us discuss the case of a standard cell complex, given by the square lattice tessellation of the $2D$ torus. In this case if one starts from a random initial conditions where the phases of the oscillators are drawn from the same distribution, one effectively considers a  perturbation that has zero projection on the neutrally stable modes of the kernel the GTS appears to be effectively asymptotically stable. However, the marginal stability of the GTS  is clearly noticeable if there is a bias in the initial distribution of the phases. For instance, when starting from an initial condition in which the phases associated with the edges in the $x$-direction have a bias with respect to the phases of the edges in the $y$-direction, the effect of this marginal stability can be clearly appreciated from the observation that GTS is not achieved as revealed by the fact that $R_2$ does not converge to one (see Figure $\ref{fig_cell0}$ ). We note, however, that edges in the $x$-direction synchronize among them-selves and that edges in the $y$-direction synchronize  among themselves. This is evident by considering the order parameter $R_{xy}$ given by
\bea
R_{xy}  = \frac{1}{N_{[n]}}\left[\left| \sum_{\alpha \in Q_x} \frac{\phi_\alpha}{\rho_0}\right| + \left| \sum_{\alpha \in Q_y} \frac{\phi_\alpha}{\rho_0}\right|\right],
\eea
where $Q_x,Q_y$ indicate respectively the set of edges in the $x$ direction and in the $y$ direction. When $R_{xy}= 1$ as in Figure $\ref{fig_cell0}$(g-i),  all the oscillators on the $x$-edges oscillate at unison, and all the oscillators on the $y$-edges oscillate at unison. However, their relative phase can be arbitrary and specifically will not be a multiple of $\pi$ if $R_2\neq 1$.  All together this discussion indicates that GTS on a $2D$ torus is marginally stable with respect to perturbations aligned along eigenvectors of type ${\bf u}$ belonging to the kernel od ${\bf L}_{[1]}$.

In the following, we will present our theoretical predictions for the occurrence of the GTS on generalized complexes.
In particular, we will show that any directed complex always admits a global synchronized state. However, this state cannot be asymptotically stable. The hollow complex representation, instead, can favor both the existence and the stability  of the GTS state with respect to undirected standard simplicial complexes. Interestingly, this benefit vanishes if one considers the more traditional tessellated representations of hollow simplices and cells.

\begin{figure}
    \centering
    \includegraphics[width=0.8\linewidth]{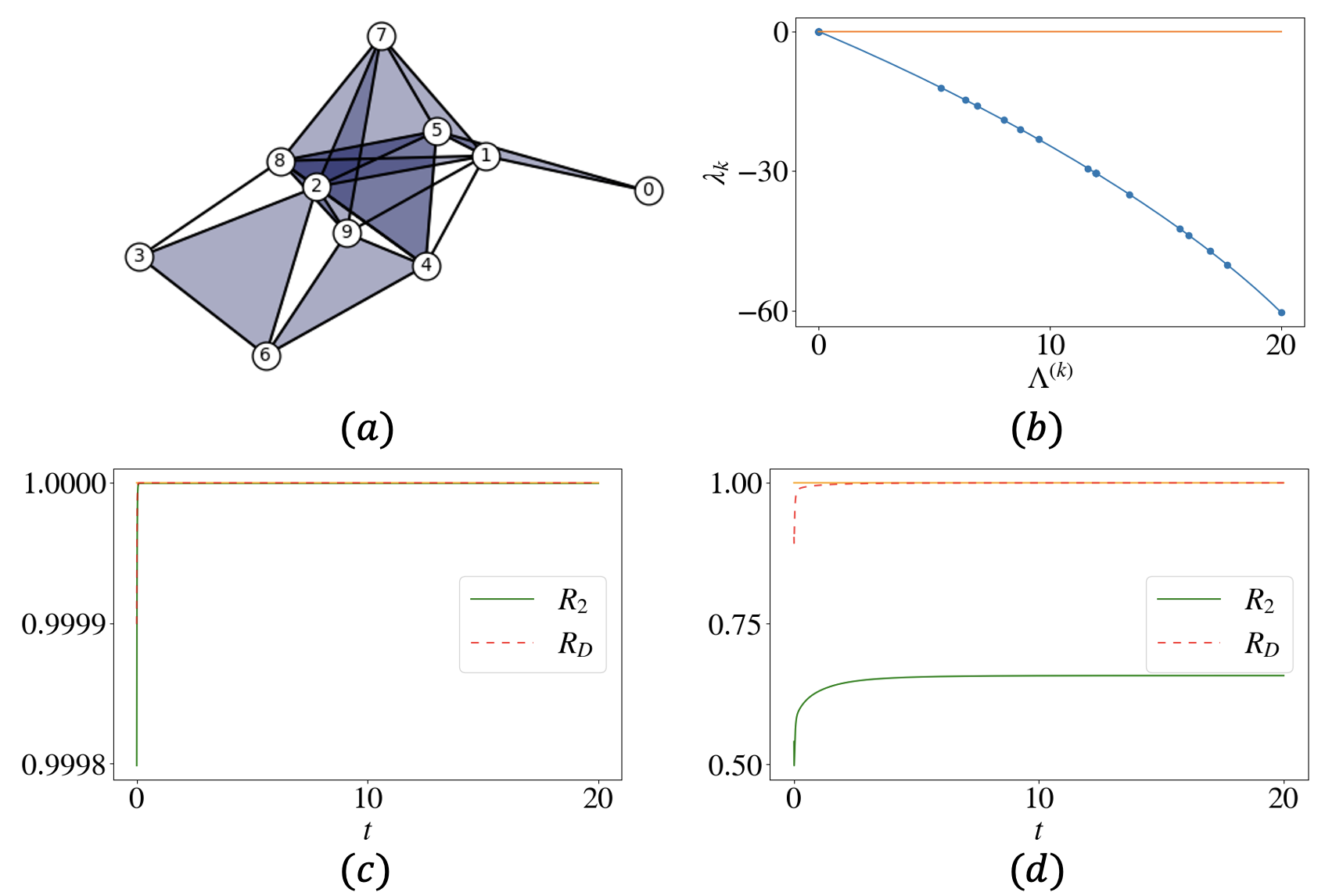}
    \caption{Illustration of the properties of the GTS of edge signals on a directed simplicial complex on which the SL dynamics is considered. Panel (a) displays a visualization of a random DSC. Panel  (b) presents the results of the  MSF approach for edge topological signal on this DSC by showing the  largest real part of the Floquet eigenvalues $\lambda_k$  associated to the $k$-th eigenmode of the Hodge Laplacian ${\bf L}_{[1]}$ with eigenvalue $\Lambda^{(k)}$. These results implies that since 
    $\lambda_k$ for $k\neq 0$ are negative, the GTS is stable with respect to perturbation orthogonal to the kernel of ${\bf L}_{[1]}$. Panel (c) demonstrates that for initial conditions that have a non-negligible projection on a single harmonic eigenvector  of type ${\bf u}$, the GTS is asymptotically stable. In this case the generalized order parameter $R_2$ convergences to  $1$.   However panel (d) demonstrates that for more general initial conditions the GTS is only neutrally stable which $R_2$ not converging to $1$ for sufficiently large times $t$. What happens in both scenarios is instead that the oscillators corresponding to the same undirected simplex, oscillate in pairs as revealed by the directed order parameter $R_D$ converging to  $R_D=1$ in both panel (c) and (d). The parameters for the SL model are $\delta=40+4.3\textrm{i},\mu=40+40.1\textrm{i},\sigma=1+1.1\textrm{i},m=3$.
    }
    \label{fig2}
\end{figure}
\subsection{Global Topological Synchronization on generalized complexes}
In this section, we investigate under which conditions GTS can be observed on generalized complexes including DSC, HSC, THSC and also DCC, HCC, THCC.
\subsubsection{GTS on DSCs and DCCs}
The  DSC and DCC provide  rather interesting generalized complexes where the GTS synchronization has non trivial properties which goe beyond what is known for global synchronization of node-based dynamics as well as for GTS on standard double orientation complexes.
First of all, here we reveal that GTS can always exist on DSC and DCC, regardless of their topology. Interestingly, for DSC and DCC the GTS exists also for generic dynamical systems of type (\ref{GTS_dyn}) where the functions $F,h$ do not need to be necessarily odd. Secondly, however, we observe  that this state is never asymptotically stable on DSC and DCC because the kernel of the generalized Hodge Laplacian ${\bf L}^{(D)}$ is highly degenerate with eigenvectors of type ${\bf u}$.
The first observation is  a direct consequence of adopting single oriented simplices (or cells). In fact, for any generic DSC (or DCC) we have always  
\bea
{\bf L}_{[n]}^{(D)}{\bf u}={\bf 0},
\eea
which clearly satisfies Eq.($\ref{gen_GTS_Laplacian})$ (see details in Appendix \ref{ApA}).
This is akin to the general scenario for node-based dynamics on graphs, which is not valid on undirected standard  complexes but always holds if the  complexes under consideration are directed.
The second observation regarding the stability of the GTS state is derived from the fact that  the harmonic eigenvalue  of ${\bf L}_{[n]}^{(D)}$ is not only highly degenerate, but the kernel of the generalized Hodge Laplacian also includes many eigenvectors of type ${\bf u}$ (see Appendix \ref{ApA}). This is a rather severe situation in the DSC and DCC, which cannot be easily overcome by considering the space orthogonal to all degenerate harmonic eigenvectors of type ${\bf u}$. Thus, although GTS exists for any arbitrary DSC and DCC, this dynamical state is not typically observed on these structures.
What happens, instead in a typical scenario, is that the phases of pairs of oscillators associated with either one of the possible single orientations of any original undirected cells, synchronize. However, different pairs of directed cells can have arbitrary difference in the phases.
This scenario can be revealed by observing $R_2\neq 1$ with $R_{D}=1$ where $R_D$ is the directed order parameter indicating synchronization of each pair of oscillators associated to the same cell. This latter order parameter is defined as 
\bea
R_D = \frac{1}{N_{[n]}\rho_0} \sum_{\alpha \in \mathcal{Q}_n^{(U)}}\left|  {\phi_\alpha}+{\phi_{-\alpha}} \right|,
\eea
with $Q_n^{(U)}$ indicating only the set of cells $\alpha$ with positive orientation, while $-\alpha$ indicates the directed cell that has the opposite orientation of $\alpha$.
The GTS on an arbitrary DSC is numerically illustrated in Figure \ref{fig2} where we simulated a higher-order topological SL model on edge signals on a random simplicial complex, by showing that  the GTS state exists but in the best dynamical scenario, it is  only neutrally stable with simplices only synchronizing in pairs as indicated by $R_D$ converging to one. The dynamical properties of GTS on DCC are completely analogous.

\begin{figure}
    \centering
    \includegraphics[width=0.8\linewidth]{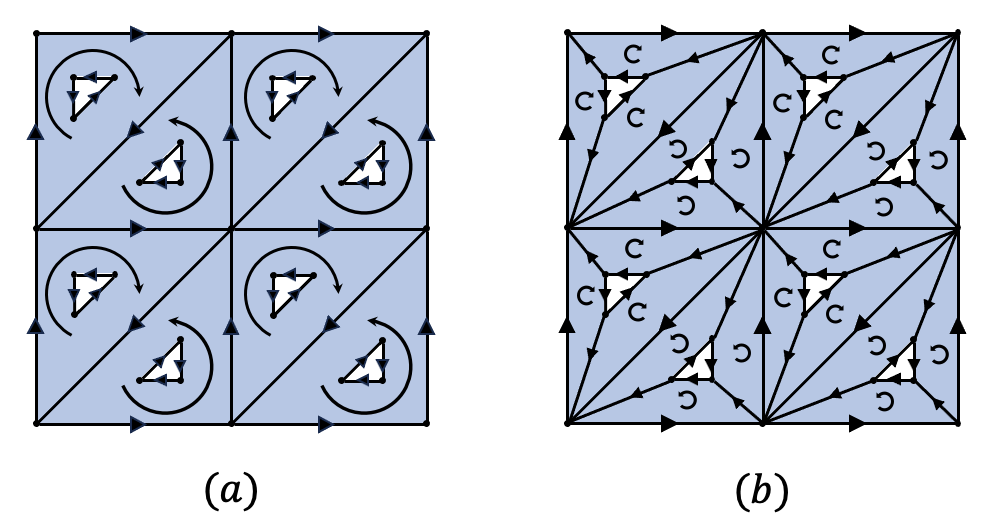}
    \caption{Illustration of the HSC and THSC constructed starting from a tessellation of a $2D$ torus. (a) The hollow triangulated torus  is a two dimensional HSC constructed from square lattice tesselation of  $2D$ torus  with periodic boundary conditions. Each  periodic unit is triangulated by two hollow 2-simplices. (b) The THSC associated to the same  triangulated torus is constructed by connecting the replica nodes of the inner triangle to the corresponding nodes on the outer triangle with edges.}
    \label{fig5}
\end{figure}

\subsubsection{GTS on HSCs and HCCs and on their associated THSCs and THCCs}
\begin{figure}
    \centering
    \includegraphics[width=0.8\linewidth]{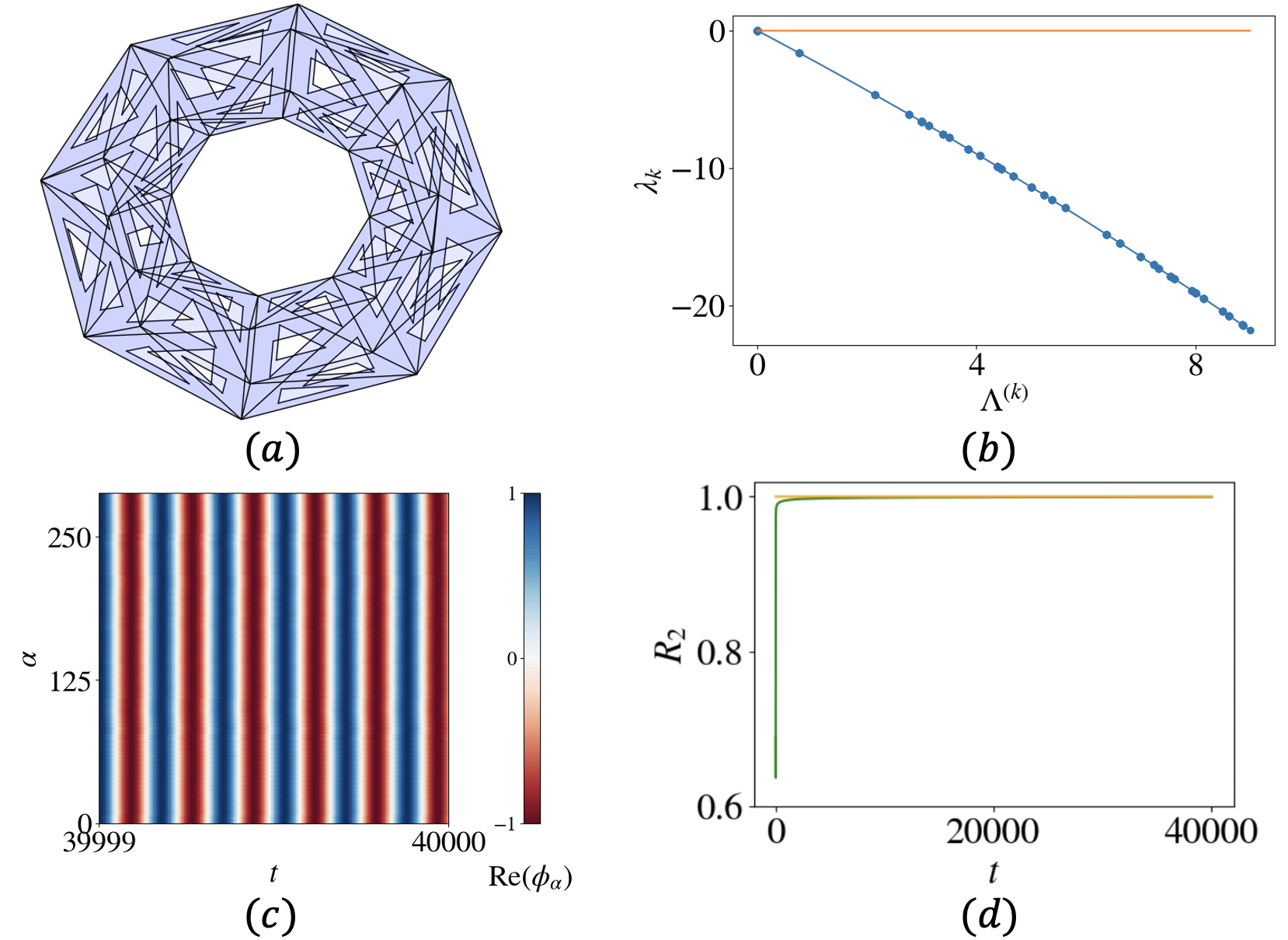}
    \caption{ 
    Illustration of the  GTS of edge signals on a hollow triangulated torus (HSC). Panel (a) shows a visualization of a hollow triangulated torus. Panel (b) provide evidence that the MSF for topological signals predict that the GTS is stable. Panel (c)  displays the temporal evolution of the real part of the topological signal $\mbox{Re}(\phi_{\alpha})$ for each edge $\alpha$ of the HSC as a function of time $t$.  Panel (d) displays the generalized order parameter $R_2$ as a function of time $t$, showing  fast convergence to $1$ indicating stable GTS.  The parameters for the SL model are $\delta =40+4.3\textrm{i},\mu=40+40.1\textrm{i},\sigma=1+1.1\textrm{i},m=3$.}
    \label{fig3}
\end{figure}

\begin{figure}
    \centering
    \includegraphics[width=0.8\linewidth]{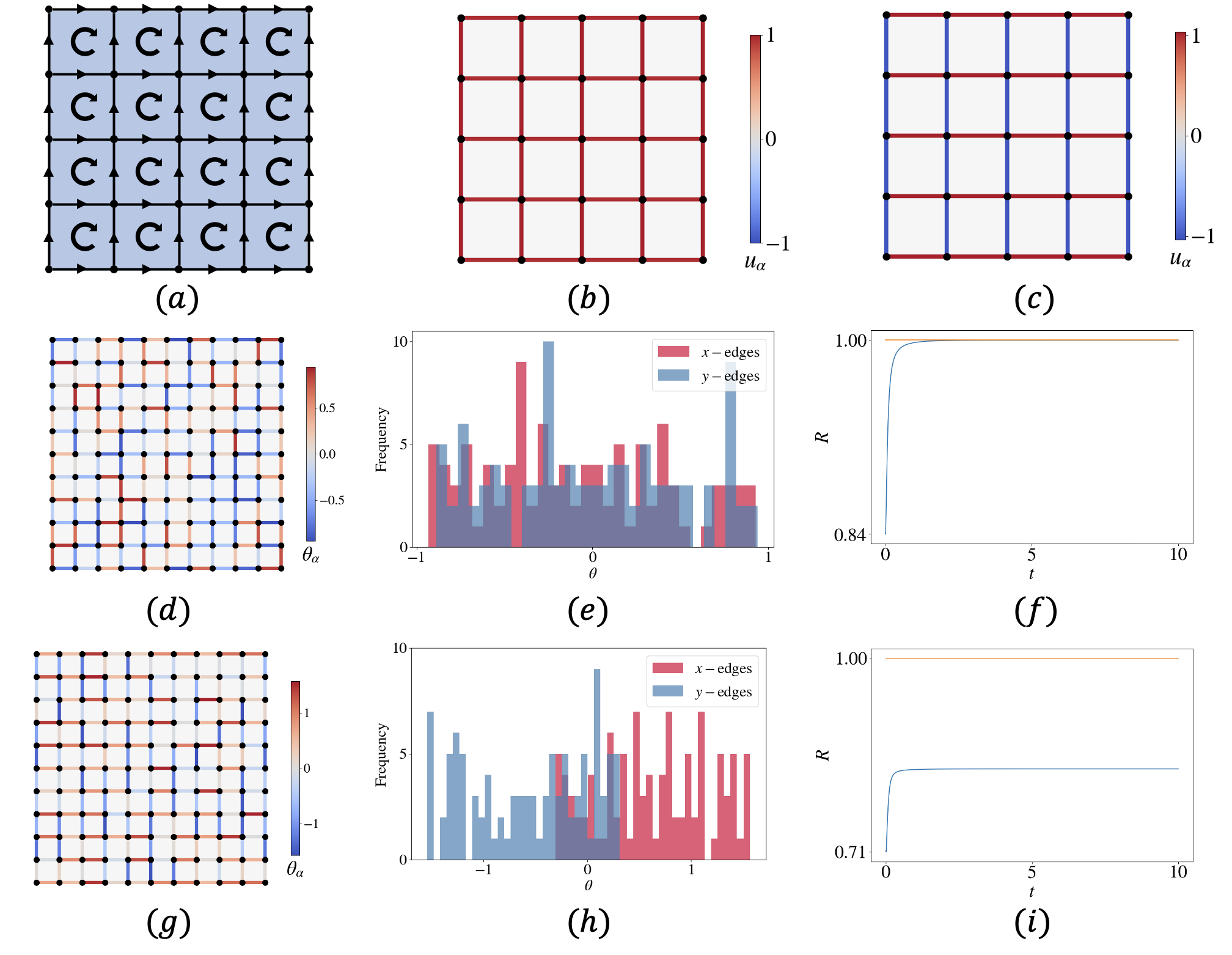}
    \caption{Illustration of   GTS on a THCC constructed from the hollow triangulated torus showing that such state  does not exists on the THCC. Panel (a) displays a visualization of a THCC constructed from the hollow triangulated torus. Panel (b) illustrates the results of the MSF which would indicate that, if the GTS were to exist, it would be stable with respect to perturbations orthogonal to the  kernel of the Hodge Laplacian ${\bf L}_{[1]}$. Panel (c) provides the representation of the temporal evolution of the real part of the edge topological signal $\mbox{Re}(\phi_{\alpha})$ associated to each edge $\alpha$ as a function of time $t$. Panel (d) displays the generalized order parameter $R(t)$ during the temporal evolution showing that the global synchronization state cannot be dynamically achieved as actually this dynamical state does not exists in this THCC. The parameters for the SL model are $\delta=40+4.3\textrm{i},\mu=40 +40\textrm{i}\sigma=1+1.1\textrm{i},m=3$.}
    \label{fig4}
\end{figure}

While DSCs and DCCs always allow  a GTS state, this state can exist only under some conditions on HSCs and HCCs. If we focus on the $(D-1)$ topological signal of a $D$ dimensional HSC, (or HCC) we can show that the first of the two conditions indicated in Eq.(\ref{gen_GTS_Boundary}), i.e.
 \bea
{\bf B}_{[D]}^{(H),\top}{\bf u}={\bf 0}
\eea
with  is always satisfied. Thus, this result indicates that a $(D-1)$-dimensional topological signal can undergo GTS if and only if
\bea
{\bf B}_{[D-1]}^{(H)}{\bf u}={\bf 0},
\label{bh}
\eea
provided  the GTS dynamics is stable.
Note, however, that the THSC (and THCC) corresponding to a given HSC (or HCC) might not sustain the GTS because Eq.(\ref{bh}) is satisfied for the HSC (or HCC) but the corresponding condition is not satisfied for the THSC, or (THCC) i.e.
\bea
{\bf B}_{[D-1]}{\bf u}\neq {\bf 0}.
\label{bthcc}
\eea
In order to demonstrate these results, we have considered a triangulated $2D$ hollow torus and defined the corresponding HSC and THSC (see Figure $\ref{fig5}$). Note that the first of these structures satisfies Eq.(\ref{bh}) while for the second complex we have Eq.(\ref{bthcc}).
In  Figure \ref{fig3} we show that  GTS is achievable for edge signals on the HSC constructed from the $2D$ torus, whereas Figure \ref{fig4} illustrates that GTS cannot be achieved on the corresponding THSC. Similar results hold for HCC and THCC.

\begin{figure}
    \centering
    \includegraphics[width=0.8\linewidth]{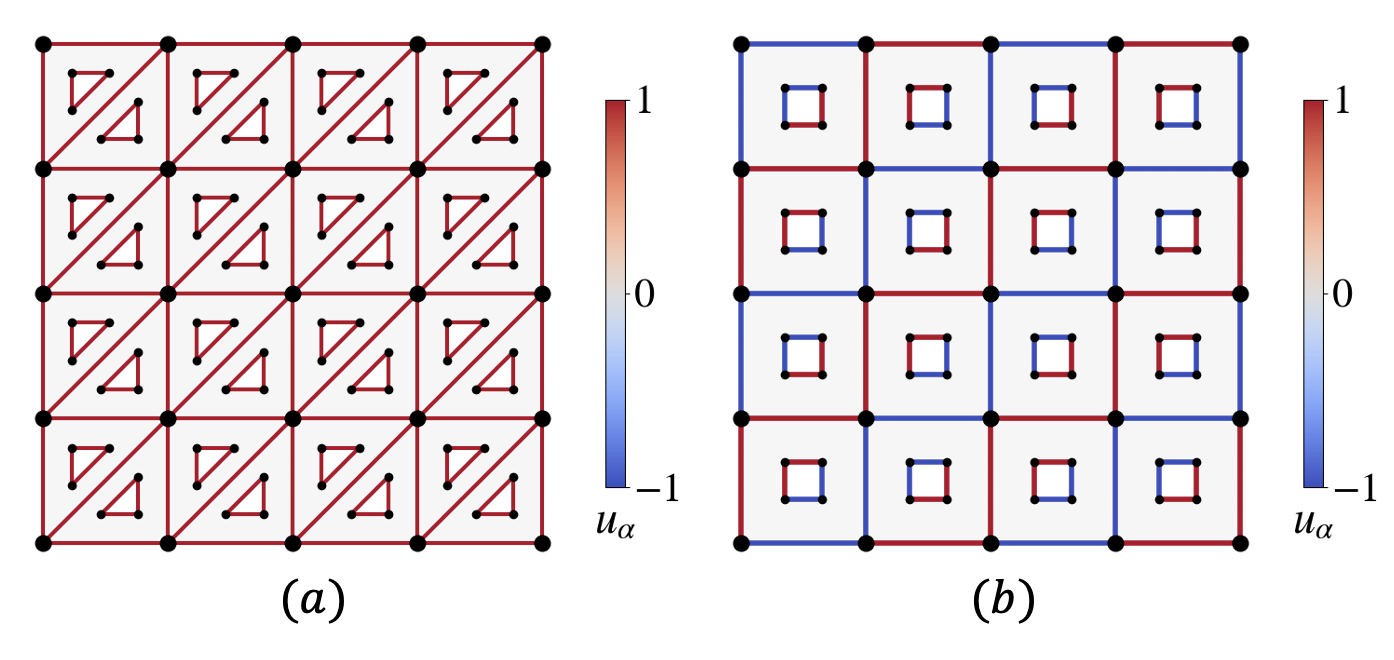}
    \caption{Visualization of the single harmonic eigenvector of type ${\bf u}$ (i.e. having elements of constant absolute value) for the $2D$ torus HSC (panel (a)) and for the $2D$ torus HCC (panel (b)).}
    \label{fig9}
\end{figure}

\begin{figure}
    \centering
    \includegraphics[width=0.8\linewidth]{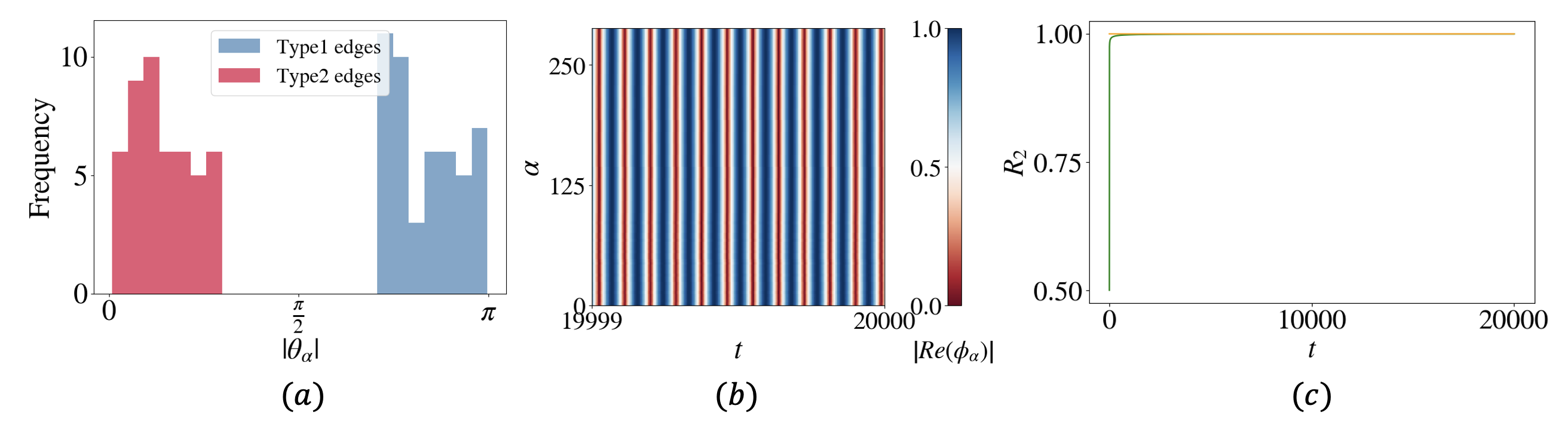}
    \caption{Illustration of GTS of edge signals on a hollow cell complex torus (HCC). Panel (a) show the distribution of the absolute values of phases $\left|\bm{\phi}_\alpha\right|$ associated to different edges classified by the red and blue colored edges indicated in Panel(b) in Figure \ref{fig9}. Panel (c) displays the temporal evolution of the real part of the absolute value of the real part of the topological signal $\left|Re(\bm{\phi}_\alpha)\right|$. Panel (c) displays the generalized order parameter $R_2$ as a function of time $t$, showing  fast convergence to $1$ indicating stable GTS.  The parameters for the SL model are $\delta =40+4.3\textrm{i},\mu=40+40.1\textrm{i},\sigma=1+1.1\textrm{i},m=3$.}
    \label{fig10}
\end{figure}

We observe that both HSC and HCC can favor the existence and the stability of GTS. Indeed, under the mentioned topological  conditions, odd-dimensional topological signals can globally synchronize, also on HSC while this is not possible in (unweighted) simplicial complexes with standard simplices. Moreover, HSC and HCC can favour the stability of the GTS as the degeneracy of harmonic eigenvectors of type ${\bf u}$ might be reduced with respect to their original simplicial or cell complex. This is illustrated in Figures \ref{fig9} and \ref{fig10} for the case of the HSC and the HCC built respectively from a triangulated $2D$ torus and a square lattice tesselation of the same structure. As we have discussed previously,  the square lattice tessellation of the torus has two eigenvectors of type ${\bf u}$ in the kernel of their Hodge Laplacian ${\bf L}_{[1]}$. Instead, there is a unique eigenvector of type ${\bf u}$ in the kernel of ${\bf L}_{[1]}$ of the triangulated HSC and  the HCC -provided that for the HCC we consider an original $2D$ torus with an even number of nodes on each side of the square lattice (see Figure \ref{fig9}).   This is a direct consequence of the requirement that these harmonic eigenvectors are divergence-free. Thus, these results demonstrate that even if in the original torus GTS is only marginally stable (\ref{fig_cell0}), on the HSC and the HCC GTS is asymptotically stable (see Figure $\ref{fig10}$).

\section{Conclusions}
This work investigates the topology and the dynamics defined on generalized complexes by including directed (DSC and DCC) and hollow (HSC and HCC) and tessellated hollow (THSC and  THCC) complexes.
By GTS here we define the dynamical state of topological signals, i.e., variables associated with each $n$-dimensional simplex of the structure, in which each dynamical variable indicates an identical oscillator oscillating in unison with all the other ones.
Specifically, here it  is shown that GTS defined on directed complexes, in which simplices have a single orientation, always admit a GTS state but that this state cannot be asymptotically stable to the degeneracy of  the kernel of their associated directed Hodge Laplacian. 
This result demonstrates that on DSC and DCC,  the topological obstruction of GTS observed on standard simplicial complexes can be completely overcome, but this come at the cost, revealed by the stability properties of this state, as GTS cannot be asymptotically stable.
When we investigate GTS on HSC we observe that more restrictions applies for the existence of the GT state, yet specific HSC can allow stable synchronization of odd-dimensional topological signals such as edge signals which instead can never be observed in standard simplicial complexes.
Interestingly, the HSCs admit a tessellated representation with standard cell complexes leading to THCC. This more traditional representation, is here shown to disfavour the presence of GTS. Specifically, we provide evidence that HSC which support GTS have a THSC representation in which GTS cannot be supported.

In summary, this work investigates how departing from the standard construction of simplicial and cell complexes by considering directed simplices with a single orientation or even hollow simplices, can affect the existence and the stability of a GTS state.  These results provide a new important perspective on the characterization of higher-order topological dynamics that might enhance our ability to characterize how the algebraic topological properties  of simplicial and cell complexes, and their associated combinatorial implications, can affect higher-order topological dynamics.

\section{Acknowledgments}
Runyue Wang acknowledges support from the China Scholarship Council (CSC).
\appendix
\section{Betti number of directed simplicial complexes}
\label{ApA}
In this Appendix, our goal is to demonstrate the relation between the topology of the DSC and the DCC and the dimension of the kernel of their associated directed Hodge Laplacians ${\bf L}_{[n]}^{(D)}$.
Specifically, we want to demonstrate Eqs. (\ref{betti_D}) that we rewrite here for convenience,
\bea
\mbox{dim}(\mbox{ker}({\bf L}_{[0]}^{(D)}))&=&\beta_{0},\nonumber \\
\mbox{dim}(\mbox{ker}({\bf L}_{[n]}^{(D)}))&=&N_{[n]}+\beta_{n},\; \; \mbox{for}\; n>0.
\label{betti}
\eea
In order to show this relation we notice that since we have 
\bea
{\bf B}_{[1]}^{(D)}&=&{\bf B}_{[1]}\mathcal{I}_{[1]},\nonumber \\
{\bf B}_{[n]}^{(D)}&=&\mathcal{I}_{[n-1]}^{\top}{\bf B}_{[n]}\mathcal{I}_{[n]}\; \; \mbox{for}\; n>1,
\eea
the directed Hodge Laplacians can be expressed as 
\bea
{\bf L}_{[0]}^{(D)}&=&2{\bf L}_{[0]},\nonumber \\
{\bf L}_{[1]}^{(D)}&=&\mathcal{I}_{[1]}^{\top}({\bf L}_{[1]}^{down}+2{\bf L}_{[1]}^{up})\mathcal{I}_{[1]}\nonumber \\
{\bf L}_{[n]}^{(D)}&=&2\mathcal{I}_{[n]}^{\top}{\bf L}_{[n]}\mathcal{I}_{[n]}\; \; \mbox{for}\; n>1,
\eea
where here we indicate with ${\bf L}_{[n]}$ the Hodge Laplacian of their corresponding undirected simplicial complexes built by  simplices having double orientation.
From the above relation, Eqs.(\ref{betti}) follow directly.
To show this, let us investigate the bilinear forms
\bea
{\bf X}^{\top}{\bf L}_{[0]}^{(D)}{\bf X}&=&2{\bf y}_{[0]}^{\top}{{\bf L}_{[0]}}{\bf y}_{[0]}\nonumber \\
{\bf X}^{\top}{\bf L}_{[1]}^{(D)}{\bf X}&=&{\bf y}_{[1]}^{\top}{({\bf L}_{[1]}^{down}+2{\bf L}_{[1]}^{up})}{\bf y}_{[1]} \nonumber\\
{\bf X}^{\top}{\bf L}_{[n]}^{(D)}{\bf X}&=&2{\bf y}_{[n]}^{\top}{\bf L}_{[n]}{\bf y}_{[n]}
\eea
where 
\bea
{\bf y}_{[0]}&=&{\bf X}\nonumber \\
{\bf y}_{[n]}&=&\mathcal{I}_{[n]}{\bf X}\;\; \mbox{for}\; n>0.
\eea
For any choice of $0\leq n\leq D$, these bilinear forms vanish for 
\bea
{\bf y}_{[n]}={\bf 0},\quad\mbox{or}\quad {\bf y}_{[n]}\in \mbox{ker}({\bf L}_{[n]}),
\label{possibilities}
\eea
where we have used Hodge decomposition implying that 
\bea
\mbox{ker}({\bf L}_{[1]})=\mbox{ker}({{\bf L}_{[1]}^{down}+2{\bf L}_{[1]}^{up})}.
\eea
For $n=0$ this occurs for a non trivial choice of ${\bf X}$ only if
\bea
{\bf X}\in \mbox{ker}({\bf L}_{[0]}).
\eea
Thus, we obtain the first of the Eqs.(\ref{betti}).
For $n>0$ the first possibility outlined by Eq.(\ref{possibilities}), ${\bf y}_{[n]}={\bf 0}$ occurs if 
\bea
{\bf X}\in \mbox{ker}({\mathcal{I}}_{[n]})=\mbox{span}\begin{pmatrix}{\bf f}\\{\bf f}\end{pmatrix}.
\eea
where ${\bf f}\in \mathbb{R}^{N_{[n]}}$ is a generic nonzero vector.
The second possibility outlined by Eq.(\ref{possibilities}), i.e. ${\bf y}_{[n]}\in \mbox{ker}({\bf L}_{[n]})$ occurs for 
\bea
{\bf X}\in(\mbox{ker}(\mathcal{I}_{[n]}))^{\perp}=\mbox{span}\begin{pmatrix}{\bf f}\\-{\bf f}\end{pmatrix}
\eea
when 
\bea
{\bf y}_{[n]}=\mathcal{I}_{[n]}{\bf X}=\mathcal{I}_{[n]}\begin{pmatrix}{\bf f}\\-{\bf f}\end{pmatrix}=2{\bf f}\in \mbox{ker}({\bf L}_{[n]})
\eea
By using 
\bea
\mbox{dim}(\mbox{ker}({\mathcal{I}}_{[n]}))=N_{[n]},\quad \mbox{dim}(\mbox{ker}({{\bf L}}_{[n]}))=\beta_{n}
\eea
we finally obtain the second of Equations (\ref{betti}), i.e.
\bea
\mbox{dim}(\mbox{ker}({\bf L}_{[n]}))=N_{[n]}+\beta_{n}.
\eea
Notice that when ${\mbox{ker}({\bf L}_{[n]})}$ does not admit an eigenvector of constant absolute value, (i.e., an eigenvector of type ${\bf u}$), the only way to obtain  harmonic eigenvectors of type ${\bf u}$ is by considering  eigenstates of type 
\bea
{\bf u}=\begin{pmatrix}{\bf f}\\{\bf f}\end{pmatrix}
\eea
where ${\bf f}$ has elements of constant absolute value. Thus these type of eigenvectors are clearly highly degenerate.

\section{Betti numbers  of hollow simplicial complexes}
\label{ApA2}
In this Appendix, our goal is to calculate the hollow Betti numbers $\beta_{[n]}^{(H)}$ determining the dimension of the kernel of the hollow Laplacians defined in Eq.(\ref{LH}), i.e., deriving the Eq.(\ref{betti_H}).
The associated hollow up- and down- Hodge Laplacian matrices ${\bf L}_{[n]}^{(H),up},{\bf L}_{[n]}^{(H),down}$ with $n< D-1$ (for the up-Hodge Laplacians) and with $n<D$ (for the down-Hodge Laplacians) are $N_{[n]}^{(H)}\times N_{[n]}^{(H)}$ matrices with a diagonal $(1+N_{[D]})\times (1+N_{[D]})$ block structure. Specifically the block structure of the down-Hodge Laplacian is given by 
\bea
{\bf L}_{[n]}^{(H),down}=\begin{pmatrix}{\bf L}_{[n]}^{down}&0 &0&\ldots &0\\
0&\hat{\bf L}_{[n]}^{down}&0&\ldots& 0\\
0&0&\hat{\bf L}_{[n]}^{down}&\ldots& 0\\
\vdots&\vdots&\vdots &\vdots&\vdots\\
0&0&0&\ldots&\hat{\bf L}_{[n]}^{down}\end{pmatrix},
\label{L_down_block}
\eea
where $\hat{\bf L}_{[n]}^{down}$ is the $n$-th down-Hodge Laplacian matrix associated with a single undirected empty $D$-simplex.
A similar expression holds for the ${\bf L}_{[n]}^{(H),up}$ Laplacians with $n< D-1$, thus the Hodge Laplacian matrix ${\bf L}_{[n]}^{(H)}$ has a diagonal block structure given by 
\bea
{\bf L}_{[n]}^{(H)}=\begin{pmatrix}{\bf L}_{[n]}&0 &0&\ldots &0\\
0&\hat{\bf L}_{[n]}&0&\ldots& 0\\
0&0&\hat{\bf L}_{[n]}&\ldots& 0\\
\vdots&\vdots&\vdots &\vdots&\vdots\\
0&0&0&\ldots&\hat{\bf L}_{[n]}\end{pmatrix},
\eea
where $\hat{\bf L}_{[n]}$ is the $n$-th  Hodge Laplacian matrix associated to a single undirected  empty $D$-simplex, i.e.
\bea
\hat{\bf L}_{[n]}^{(r)}={\bf B}_{[n]}^{(r),\top}{\bf B}_{[n]}^{(r)}.
\eea
with ${\bf B}_{[D]}^{(r)}$ defined in Eq.(\ref{boundary_r}).
From this expression, by taking into account that the Hodge Laplacian $\hat{\bf L}_{[n]}$ has a kernel of zero dimension for $0<n<D-1$ and equal to one for $n=0$,
it follows directly that for $n<D-1$
\bea
\mbox{dim}\ (\mbox{ker}({\bf L}_{[n]}^{(H)}))\equiv\beta_n^{(H)}=\beta_n+\delta_{n,0}N_{[D]},
\eea
where $\delta_{x,y}$ indicates the Kronecker delta.

For $n=D$ the Hodge Laplacian is given by
\bea
{\bf L}_{[D]}^{(H)}={\bf B}_{[D]}^{(H),\top}{\bf B}_{[D]}^{(H)}
\eea
which can be equivalently expressed as
\bea
{\bf L}_{[D]}^{(H)}={\bf L}_{[D]}+\sum_{r=0}^{N_{[D]}}\hat{\bf L}_{[D]}^{(r)}.\eea

We note that ${\bf L}_{[D]}$ is semipositive definite but $\hat{\bf L}_{[D]}^{(r)}$ is the $D$-Hodge Laplacians associated to the single (full) $D$ simplex $\alpha_r$, thus ${\bf L}_{[D]}^{(H)}$ is positive definite.
It follows that 
\bea
\mbox{dim}\ (\mbox{ker}({\bf L}_{[D]}^{(H)}))\equiv\beta_D^{(H)}=0.
\label{betti_HDA}
\eea
It remains to calculate the dimension of the kernel of the $(D-1)$ hollow Hodge Laplacian. Since the hollow Hodge Laplacian obeys Hodge decomposition, it follows that the 
\bea
\mbox{dim}\ (\mbox{ker}({\bf L}_{[D-1]}^{(H)}))&=&N^{(H)}_{[D-1]}- \mbox{rank}({\bf L}_{[D-1]}^{(H),down})-\mbox{rank}({\bf L}_{[D-1]}^{(H),up})\nonumber \\
&=&\mbox{dim}\ (\mbox{ker}({\bf L}_{[D-1]}^{(H),down}))-\mbox{rank}({\bf L}_{[D-1]}^{(H),up}).
\label{HodgeD_HD}
\eea
Since the hollow Laplacians ${\bf L}_{[D-1]}^{(H),up}$ and ${\bf L}_{[D]}^{(H),down}$ share the same non-zero spectrum, 
by using Eq.(\ref{betti_HDA}) it follows that 
\bea
\mbox{rank}({\bf L}_{[D-1]}^{(H),up})=N^{(H)}_{[D]}=N_{[D]}.
\label{betti_HD2}
\eea
Moreover, since ${\bf L}_{[D-1]}^{(H),down}$ has the diagonal block structure indicated by Eq.(\ref{L_down_block}), it is immediate to show that its kernel has dimension
\bea
\mbox{dim}\ (\mbox{ker}({\bf L}_{[D-1]}^{(H)}))=\mbox{dim}\ (\mbox{ker}({\bf L}_{[D-1]}^{down}))+N_{[D]}\mbox{dim}\ (\mbox{ker}(\hat{\bf L}_{[D-1]})).\label{kDm1}
\eea
The dimension of the kernel of $\hat{\bf L}_{[D-1]}$ is given by one, as this Hodge Laplacian is the one of an empty $D$-dimensional simplex, i.e., 
\bea
\mbox{dim}\ (\mbox{ker}(\hat{\bf L}_{[D-1]}))&=&1.
\label{Ldm11}
\eea
The dimension of the kernel of the ${\bf L}_{[D-1]}^{down}$ is given by
\bea
\mbox{dim}\ (\mbox{ker}({\bf L}_{[D-1]}^{down}))&=&\beta_{D-1}-\beta_D+N_{[D]}.\label{LDm1d}
\eea
To prove this result, we first notice, that due to Hodge decomposition, we have that
\bea
\mbox{dim}\ (\mbox{ker}({\bf L}_{[D-1]}))=\beta_{D-1}&=&N^{(H)}_{[D-1]}-\mbox{rank}({\bf L}_{[D-1]}^{down})-\mbox{rank}({\bf L}_{[D-1]}^{up}),
\eea
This implies that 
\bea
\mbox{dim}\ (\mbox{ker}({\bf L}_{[D-1]}^{down}))=N^{(H)}_{[D-1]}-\mbox{rank}({\bf L}_{[D-1]}^{down})=\beta_{D-1}+\mbox{rank}({\bf L}_{[D-1]}^{up}).
\eea
By using the fact that the dimension of the rank of ${\bf L}_{[D-1]}^{up}$ is the same as the dimension of the rank of ${\bf L}_{[D]}^{down}={\bf L}_{[D]}$ we have 
\bea
\mbox{rank}({\bf L}_{[D-1]}^{up})=\mbox{rank}({\bf L}_{[D]})=N_{[D]}-\beta_D,
\eea
we can finally obtain Eq.(\ref{LDm1d}).
Finally, using Eq.(\ref{kDm1}), together with Eq.(\ref{Ldm11}) and (\ref{LDm1d}) and inserting this in Eq.(\ref{HodgeD_HD}),
 it follows that the rank of ${\bf L}_{[D-1]}^{(H),down}$ is given by
\bea
\mbox{dim}\ (\mbox{ker}({\bf L}_{[D-1]}^{(H)}))\equiv\beta_{D-1}^{(H)}=\beta_{D-1}-\beta_D+N_{[D]}.
\label{betti_HD3}
\eea

\section{MSF of the Stuart-Landau model }
\label{ApB}
In this Appendix, we derive the MSF establishing the dynamical conditions that ensure the stability of the synchronized dynamics of the Stuart-Landau (SL)  model in which all the topological signals follow the same limit cycle solution $\hat{z}(t)$ indicated in Eq.(\ref{z})  with respect to perturbations orthogonal to the kernel of the generalized Hodge Laplacian. 
Note that the  SL is generic being the normal form of the supercritical Hopf bifurcation, but at the same time it allows a simple analysis being the linear system autonomous and thus it is enough to compute its spectrum.

We recall that the MSF can be applied only if the GTS state exists, and only in this condition it is useful to determine the values of the dynamical parameters that ensure its stability with respect to perturbations orthogonal to the kernel of the generalized Hodge Laplacian. 
In order to treat deviations  away from the GTS state we introduce the small quantities $r_{\alpha}(t)$ and $\theta_{\alpha}(t)$ which determine the small deviations of $\phi_{\alpha}$ from $\hat{z}(t)$, i.e.
\begin{equation}
\label{eq:wpert}
\phi_{\alpha}(t)=\hat{z}(t)(1+r_{\alpha}(t))e^{i\theta_{\alpha}(t)}\, ,
\end{equation}
where $r_{\alpha}(t)$ and $\theta_{\alpha}(t)$ are real valued functions.

By using the expression for $\hat{z}(t)$ given by Eq.(\ref{z}) and inserting this expression in Eq.(\ref{SL}) we find, in first order approximation on $r_{\alpha}$ and $\theta_{\alpha}$, 
\bea
\label{eq:maineqSLlin}
\frac{dr_{\alpha}}{dt} &=& -2\delta_\Re r_{\alpha}  -\left(\frac{\delta_\Re}{\mu_{\Re}}\right)^{\frac{m-1}{2}}\sum_{{\alpha^{\prime}}=1}^{N_{[n]}} L_{[n]}^{(G)}(j,{\alpha^{\prime}})\left(m\sigma_\Re r_{\alpha^{\prime}}-\sigma_\Im \theta_{\alpha^{\prime}}\right)\\
\frac{d\theta_{\alpha}}{dt} &=& -2\mu_\Im\frac{\delta_\Re}{\mu_{\Re}} r_{\alpha}  -\left(\frac{\delta_\Re}{\mu_{\Re}}\right)^{\frac{m-1}{2}}\sum_{{\alpha^{\prime}}=1}^{N_{[n]}} L_{[n]}^{(G)}(j,{\alpha^{\prime}})\left(m\sigma_\Im r_{\alpha^{\prime}}+\sigma_\Re \theta_{{\alpha^{\prime}}}\right)\, .
\eea
By projecting $r_{\alpha}(t)$ and $\theta_{\alpha}(t)$ on the orthonormal eigenbasis $\pmb{\psi}_{[n]}^{(k)}$, $k=1,\dots,N_{[n]}$, of the generalized Laplacian matrix $\mathbf{L}_{[n]}^{(G)}$, where $G\in\{S,H,T\}$, we obtain
\begin{equation}
\label{eq:projection}
 r_{\alpha}=\sum_k \hat{r}_k \psi^{(k)}_{[n]}(\alpha) \text{ and } \theta_{\alpha}=\sum_k \hat{\theta}_k \psi^{(k)}_{[n]}(\alpha) \, ,
\end{equation}
where $\hat{r}_k$ and $\hat{\theta}_k$ obey, at the linear order,  the dynamical equations
\bea
\label{eq:maineqSLlinMSFApp}
\frac{d{\hat{r}}_k}{dt} &=& -2\delta_\Re \hat{r}_k  -\left(\frac{\delta_\Re}{\mu_{\Re}}\right)^{\frac{m-1}{2}}\Lambda^{(k)}_{[n]}\left(m\sigma_\Re \hat{r}_k-\mu_\Im \hat{\theta}_k\right)\\
\frac{d{\hat{\theta}}_k}{dt} &=& -2\mu_\Im\frac{\delta_\Re}{\mu_{\Re}} \hat{r}_k  -\left(\frac{\delta_\Re}{\mu_{\Re}}\right)^{\frac{m-1}{2}}\Lambda^{(k)}_{[n]}\left(m\sigma_\Im \hat{r}_k+\sigma_\Re \hat{\theta}_k\right)\, ,
\eea
where $\Lambda_{[n]}^{(k)}$ indicates the $k$-th eigenvalue of the generalized Laplacian matrix ${\bf L}_{[n]}^{(G)}.$

For the SL model, the MSF resorts to Floquet analysis to study the stability of the perturbations associated to the $k$-th eigenmode and imposes that the largest real part of the Floquet eigenvalues $\lambda_k=\lambda_k(\Lambda_{[n]}^{(k)})$ is negative, $\Re\lambda_k<0$, then the reference solution is stable under perturbations orthogonal to the kernel of the generalized Hodge Laplacian. 
\section*{Code Availability}
All the codes used in this work are available upon request.

\bibliographystyle{unsrt}
\bibliography{bibliography}

\end{document}